\documentclass[10pt,journal,compsoc]{IEEEtran}
\usepackage{array}
\usepackage{enumitem}
\usepackage{multirow}
\usepackage{makecell}
\usepackage[table]{xcolor}%
\usepackage[font=small,skip=0pt]{caption}
\usepackage{subfig}

\ifCLASSOPTIONcompsoc
  \usepackage[nocompress]{cite}
\else
  \usepackage{cite}
\fi
\ifCLASSINFOpdf
   \usepackage[pdftex]{graphicx}
\else
\fi
\begin{document}
%
\title{Insights from  BB-MAS - A Large Dataset for Typing, Gait and Swipes  of the Same Person on Desktop, Tablet and Phone}
%
%
%
%

\author{
\IEEEauthorblockN{Amith K. Belman\IEEEauthorrefmark{1},
Li Wang\IEEEauthorrefmark{2},
S. S. Iyengar\IEEEauthorrefmark{2}, 
Pawe\l{} \'{S}niata\l{}a\IEEEauthorrefmark{3},\\
Robert Wright\IEEEauthorrefmark{4},
Robert Dora\IEEEauthorrefmark{4},  
Jacob Baldwin\IEEEauthorrefmark{4}, Zhanpeng Jin\IEEEauthorrefmark{6} and Vir V. Phoha\IEEEauthorrefmark{1}}

\IEEEauthorblockA{\IEEEauthorrefmark{1}Syracuse University, \{akamathb,vvphoha\}@syr.edu}

\IEEEauthorblockA{\IEEEauthorrefmark{2}Florida International University, \{lwang059,iyengar\}@cs.fiu.edu}

\IEEEauthorblockA{\IEEEauthorrefmark{3}Pozna\'{n} University of Technology, pawel.sniatala@put.poznan.pl}

\IEEEauthorblockA{\IEEEauthorrefmark{4}Assured Information Security, Inc., \{wrightr,dorar,baldwinj\}@ainfosec.com}

\IEEEauthorblockA{\IEEEauthorrefmark{6}University at Buffalo, SUNY, zjin@buffalo.edu}
}

\IEEEtitleabstractindextext{%
\begin{abstract}
Behavioral biometrics are key components in the landscape of research in continuous and active user authentication. However, there is a lack of large datasets with multiple activities, such as typing, gait and swipe performed by the same person. Furthermore, large datasets with multiple activities performed on multiple devices by the same person are non-existent. The difficulties of procuring devices, participants, designing protocol, secure storage and on-field hindrances may have contributed to this scarcity. The availability of such a dataset is crucial to forward the research in behavioral biometrics as usage of multiple devices by a person is common nowadays. Through this paper, we share our dataset, the details of its collection, features for each modality and our findings of how keystroke features vary across devices. We have collected data from 117 subjects for typing (both fixed and free text), gait (walking, upstairs and downstairs) and touch on Desktop, Tablet and Phone. The dataset consists a total of about: 3.5 million keystroke events; 57.1 million data-points for accelerometer and gyroscope each; 1.7 million data-points for swipes; and enables future research to explore previously unexplored directions in inter-device and inter-modality biometrics.  Our analysis on keystrokes reveals that in most cases, keyhold times are smaller but inter-key latencies are larger,  on hand-held devices when compared to desktop. We also present; detailed comparison with related datasets; possible research directions with the dataset; and lessons learnt from the data collection.
\end{abstract}

\begin{IEEEkeywords}
Biometric Dataset, Keystroke, Gait, Stairs, Phone, Tablet, Desktop.
\end{IEEEkeywords}}

\maketitle

\IEEEdisplaynontitleabstractindextext

%
\IEEEpeerreviewmaketitle

\IEEEraisesectionheading{\section{Introduction}\label{sec:introduction}}

%
%
%
%
\IEEEPARstart{T}{he} recent advancements in active or continuous authentication using behavioral biometrics show that they are promising complements to existing one-time authentication methods like PINs, passwords, fingerprint, etc., generally deployed at the entry points of a system. Behavioral biometrics have an implicit usability advantage as the process of authentication is not separate from whatever the user intends to do with the system. In other words, whatever the user activity is, it becomes a mode of authentication. Researchers have explored various modalities such as keystrokes (\hspace{1sp}\cite{disckeyfea,keystrokemon,scankeystroke}), gait (\hspace{1sp}\cite{biogait,gaitwearable,gaitsurvey}), swipes on touch screen (\hspace{1sp}\cite{touchgest,swipegest,toucheval}) to name a few. With growing number of devices used by a person, research in continuous authentication  or behavior analysis will have span across devices and activities to stay relevant. However, the scarcity of benchmark datasets for such scenarios are a hindrance. Several attempts have been made to provide benchmark datasets for a single activity like keystrokes (\hspace{1sp}\cite{cmuks,sharedks,laserks,androidks,pressureks,prosodyks,videoks}), gait (\hspace{1sp}\cite{gait1,gait2,gait3,hugadb})  or swipes \hspace{1sp} (\hspace{1sp}\cite{toucheval,touchalytics,touch3,touchgest}) on a single device family like desktop or phone. Few attempts were also made to share benchmark datasets with multiple activities using single device (\hspace{1sp}\cite{unimibshar,har}). However, no benchmark dataset exists for multi-activity in multi-device scenario, where the same activities were performed by the same users on multiple devices. We attempt to fill this gap and provide a benchmark dataset with BB-MAS (Behavioral Biometrics Multi-device and multi-Activity data from Same users)  dataset where the same participants have provided typing, gait and swiping data on desktop, phone and tablet.

A total of 117 participants have provided data voluntarily. Each participant has performed typing (including transcription and free text), gait (including walking on a flat corridor, upstairs and downstairs) and swiping using desktop, phone  and tablet. The data collection spanned about 3 months and various anonymized demographics information is provided for each participant. The unique ID allocated to the participant is used on all devices and activities.

\subsection{Key Contributions:}
\begin{itemize} 
    \item Insights from analysis on how the keystroke feature values vary between desktop, tablet and phone for the same users are presented. We find the keyhold times are smaller and inter-key latencies are larger on hand-held devices when compared to desktop. We posit that, difference in number of contact points while typing (fewer on hand-held device) may lead to such patterns.

    \item To the best of our knowledge, a dataset with the typing, gait and touch data from the same users on desktop, tablet and phone is not available publicly at the time of this writing. With data from 117 participants our dataset stands out as unique and rich for exploration in various directions. Each participant's session ranged between 2 to 2.5 hours, resulting in a total of about: 3.5 million keystroke events; 57.1 million data-points for accelerometer and gyroscope each; 1.7 million data-points for swipes.
    
    \item Features that are commonly described in literature, for data from all activities are described, extracted and shared alongside the raw data. 
    
    \item We compare our dataset with other related datasets for keystroke, gait and swipe and highlight their novelty, differences and advantages.
    
    \item Possible research directions using the BB-MAS dataset are discussed and  lessons learnt from this elaborate data-collection effort are shared to help future researchers on similar endeavours.

\end{itemize} 

Data collection was carried out between April and June of 2017, after the IRB approval from our university. All participants signed consent forms and have willingly participated in this data collection. All data has been anonymized and any personal identifiers in the data are removed. All subjects, their data and demographic information can only be referenced through the unique participant ID provided to them.


\section{Data Collection}\label{datacollection}
The dataset was designed to capture the behavior of the same users performing various day-to-day activities, such as typing, gait and swipes on three commonly used devices such as, desktop, tablet and phone. Activities were deliberately designed to mimic real-life scenarios, for instance, the typing activity consists of both fixed and free text data, the gait activity consists of walking on flat corridors, walking downstairs and upstairs and touch and swipe data consists of activities such as reading and scrolling. The raw data and the features extracted are shared publicly and can be accessed at \textit{http://dx.doi.org/10.21227/rpaz-0h66} . 
\subsection{Devices}\label{devicesandsensors}
Three most commonly used devices type in current times were selected for our data collection. A desktop, tablet and  phone would cover most of our modern-day interactions with devices. The details of the devices used in our data collection are as below: 
\begin{itemize}
    \item \textbf{Desktops}: Two identical desktop stations were setup. Each desktop station consisted of a standard QWERTY keyboard (Dell kb212-b), an optical mouse (Dell ms111-p) and a Dell 21 inch monitor. The keystrokes, mouse movements and clicks were logged.
    
    \item \textbf{Tablets}: HTC-Nexus-9 tablets were used for the tablet section of the data collection. These tablets had a screen size of 8.9 inches, screen resolution of 1536 x 2048 pixels, device dimensions of 9 x 6 x 0.3 inches (Length X Width X Height) and weighed about 435 grams. Keystrokes, accelerometer, gyroscope and touch were logged.
    
    \item \textbf{Phones}: Two different models of  phones, Samsung-S6 and HTC-One phones were used in the data collection. The Samsung Galaxy S6 had a screen size of 5.1 inches and screen resolution of 1440 x 2560 pixels with body dimensions of 143.4 x 70.5 x 6.8 mm and weighing 138 grams, whereas the HTC-One had a screen size of 5.0 inches and  screen resolution of  1080 x 1920 pixels with body dimensions of 146.4 x 70.6 x 9.4 mm and weighing 160 grams. Keystrokes, accelerometer, gyroscope and touch were logged. The raw data files from different models are identified by the suffix in the file names explained in detail in Section \ref{sub:dataformat}.
\end{itemize}

\begin{figure}[h]
  \centering
  \includegraphics[width=0.4\linewidth]{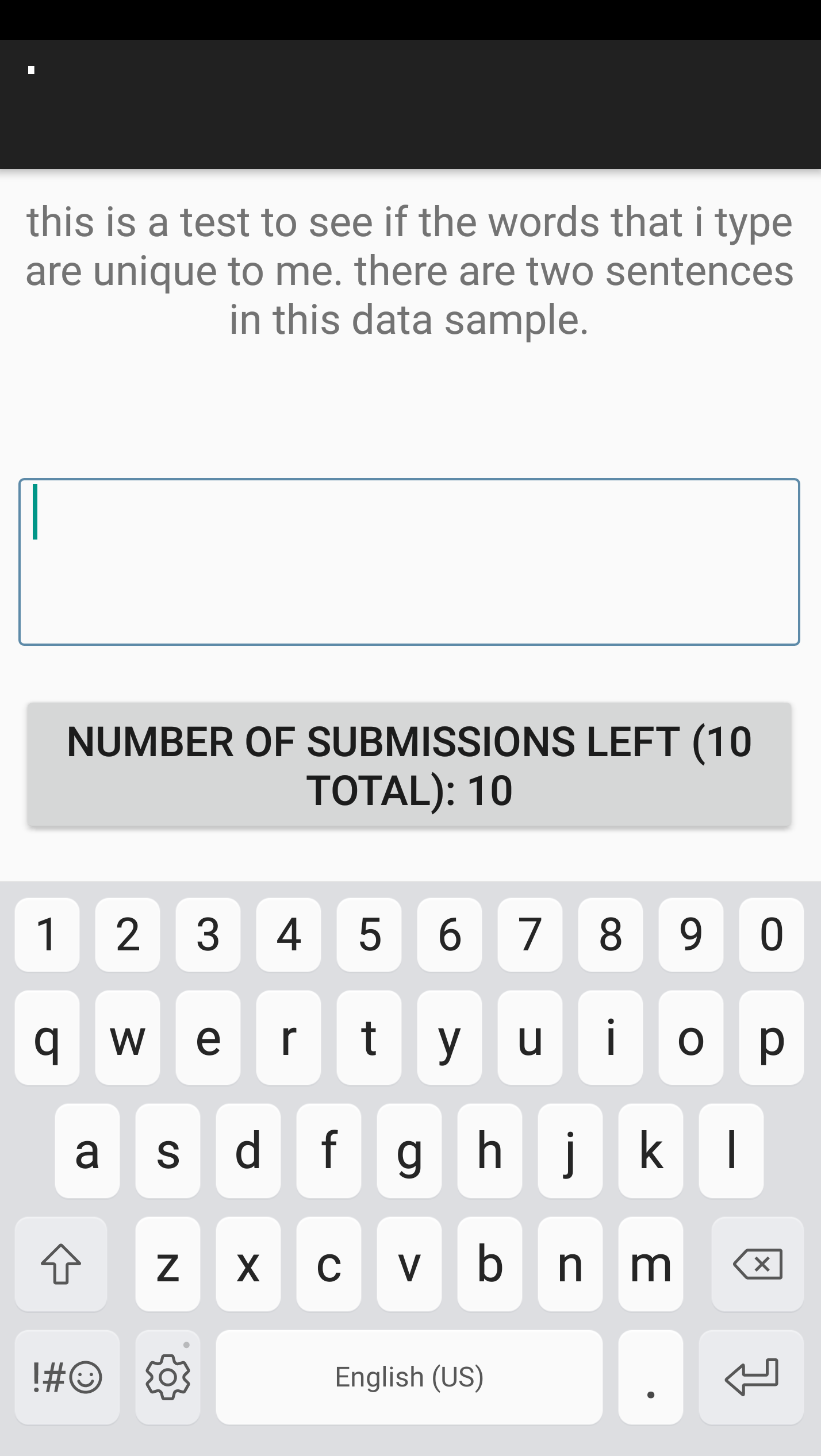}
  \caption{A screenshot from our phone application showing our keyboard to match the default android keyboard.}~\label{fig:shot}
\end{figure}

As the default android keyboard does not allow logging of keystrokes, we created and used an android qwerty keyboard on screen which was similar to the default android qwerty keyboard. The phones and tablets were locked in portrait orientation and users were allowed to type on them with any comfortable posture that they preferred. The details of the data collected from these devices and their formats is described in Section \ref{sub:dataformat}. Figure \ref{fig:shot} shows a screenshot of the application with the keyboard for phone. The application on tablet had the same layout but was scaled to match the default keyboard of an android tablet.

\begin{figure*}[h]
  \centering
  \includegraphics[width=0.8\linewidth]{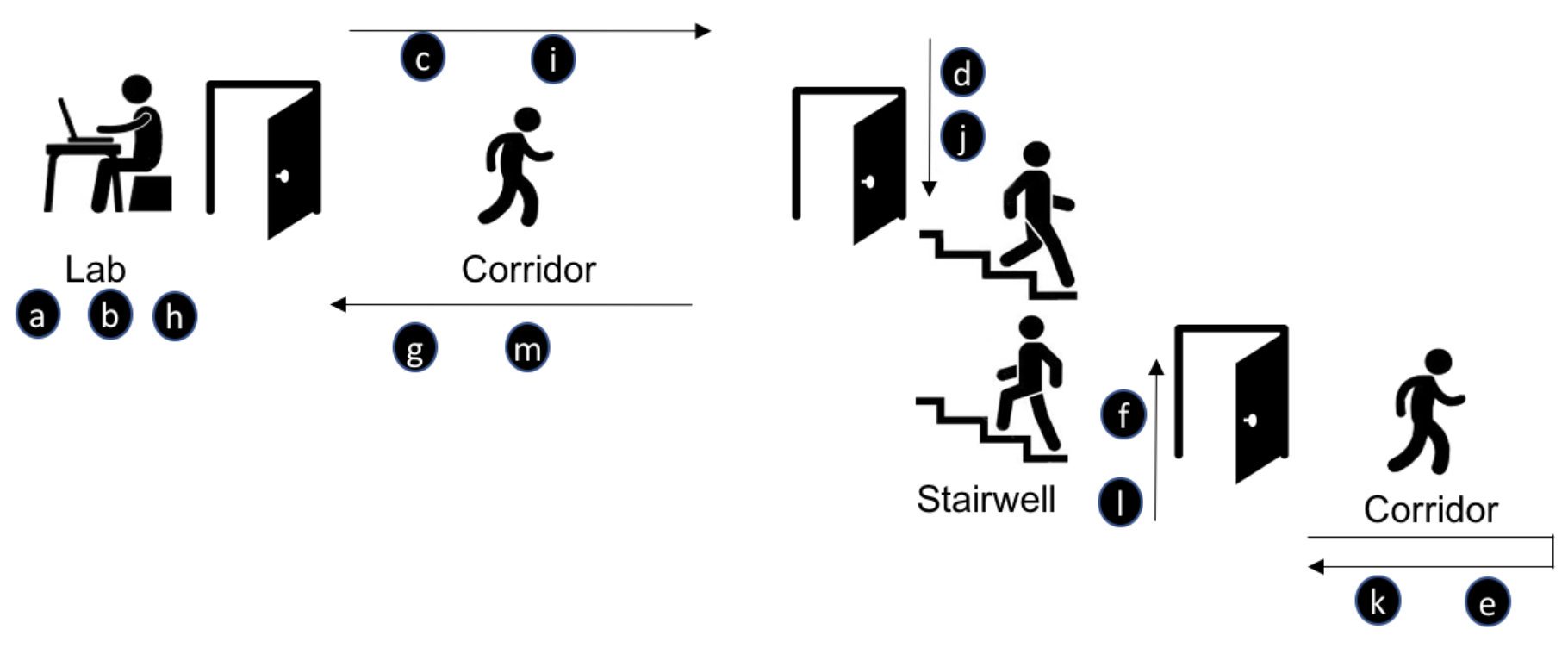}
  \caption{The data collection procedure. Tasks \textbf{a} to \textbf{m} were performed by participants in sequence, the corresponding activities and data collected are described in Table \ref{tab:CollectionTable}}~\label{fig:Collection}
\end{figure*}

\subsection{ How was the data collected? }\label{howcollected}

 Emails were sent out to all students, faculty, and staff to procure the participant population. Each participant had to spend two hours on average to perform the set of sequential tasks as illustrated in Fig. \ref{fig:Collection}.  

Upon arrival at the data collection location, each participant answered a set of questions that pertained to his/her demographics and technology usage. The participant was then assigned a unique ID and four devices: a desktop, a tablet and two phones (See Table \ref{tab:CollectionTable}). The participant then performed the tasks \textbf{a} to \textbf{m} in sequence. \textbf{a}) The participant was asked to sit at the desktop and type two sections of text (fixed-text), ten times each. Each piece of text consisted of two sentences and had an average of 112 characters. The participant was then given a shopping list consisting of six items. They had to use a popular web-browser (Mozilla Firefox) to browse for the best prices for the six items on the list while making notes (on any familiar text editor) about prices, opinions and thoughts. The participant was then given a list of 12 questions of varying cognitive loads (see Appendix \ref{appendix:cognitive} - \ref{appendix:tablet}) and asked to type their answers in any order he/she preferred for roughly about fifteen minutes. For the entire duration of task \textbf{a}, keystroke and mouse loggers were deployed on the desktop to log all the actions that the participant performed during this task. \textbf{b}) After the completion of task \textbf{a}, the participant was handed a tablet which was running an application where he/she was asked to type the two pieces of static text again followed by a series of ten questions with varying cognitive loads to be answered with a minimum of 50 characters. The questions were placed in a manner that required the participant to swipe vertically and horizontally between questions. For the entire duration of task \textbf{b}, keystroke, touch, accelerometer, and gyroscope loggers were deployed on the tablet to log all typing, swiping, touch, and movement events. 
After the completion of task \textbf{b}, the participant was asked to place a phone (Phone1) in his/her pants pocket and made to walk in a predefined path while holding the tablet in hand. The path consisted of three doorways and a stairwell, as shown in Figure \ref{fig:Collection}. The tablet displayed buttons to be pressed by the participant before and after passing through a doorway and also before and after taking the staircase. The tasks \textbf{c}, \textbf{e}, and \textbf{g}   required the participant to walk, and tasks \textbf{d} and \textbf{f} required the participant to climb downstairs and upstairs respectively. Throughout the tasks \textbf{c} to \textbf{g}, the tablet and the phone (Phone1) logged the accelerometer and gyroscope values. The tablet also logged the pressing of the buttons (doorway and staircase) by the participant.

Upon completion of task \textbf{g},  the tablet was taken from the participant and another phone (Phone2) was handed to them. For task \textbf{h}, Phone2 ran the same application as the tablet in task \textbf{b}, where the participant had to  type the two pieces of static text followed by a series of ten questions (not repeated from task \textbf{b}) with varying cognitive loads to be answered with a minimum of 50 characters, requiring the user to swipe between questions.  Phone2 logged all keystroke, touch, accelerometer and gyroscope values for typing, swiping, touch, and movement events. Tasks \textbf{i} to \textbf{m} are similar to tasks \textbf{c} to \textbf{g}, differing only in that the participant held Phone2 (instead of the tablet) and Phone1 remained in pocket while performing tasks \textbf{i} to \textbf{m}. Phone1 and Phone2 logged all accelerometer and gyroscope values. Phone2 also logged the pressing of buttons (doorway and staircase) by the participant.

\begin{table*}[h]
\caption{Data collection tasks performed by the participants. For each participant we recorded activities on four devices - a Desktop, a Tablet and two Phones (pocket and hand). }
\label{tab:CollectionTable}
  \centering
  \begin{tabular}{ccccc}
\hline\hline
Task  & Device & Activity & Data & Duration (Approx.) \\ [2pt]
\hline
&&&&\\
a & Desktop & Typing, Browsing & Keystroke and Mouse & 50 min  \\ [2pt] \hline
&&&&\\
b & Tablet & Typing &  Keystroke, Swipe, Accelerometer, Gyroscope & 25 min \\  [2pt]\hline
&&&&\\
c & \multirow{5}{*} {\makecell{Tablet (in hand) \\ Phone1 (in pocket)}} & Walking &  \multirow{5}{*}{Accelerometer, Gyroscope}  & \multirow{5}{*} {5 min}  \\  [2pt]
d & & Climbing down stairs & & \\  [2pt]
e & & Walking & &  \\  [2pt]
f & & Climbing upstairs & & \\ [2pt]
g & &Walking & & \\ [4pt] \hline
&&&&\\
h & Phone2 (in hand) & Typing & Keystroke, Swipe, Accelerometer, Gyroscope & 25 min  \\  [2pt]\hline
&&&&\\
i &\multirow{5}{*} {\makecell{Phone2 (in hand) \\ Phone1 (in pocket)}} &  Walking &  \multirowcell{5}{Accelerometer, Gyroscope} & \multirow{5}{*} {5 min} \\  [2pt]
j & & Climbing down stairs & &  \\ [2pt]
k & &  Walking & & \\  [2pt]
l & & Climbing upstairs & & \\  [4pt]
m & & Walking & & \\  [4pt]
\hline
\end{tabular}
\end{table*}

\subsection{ What is the format of the data?}\label{sub:dataformat}
\begin{figure*}
  \centering
  \includegraphics[width=0.9\linewidth]{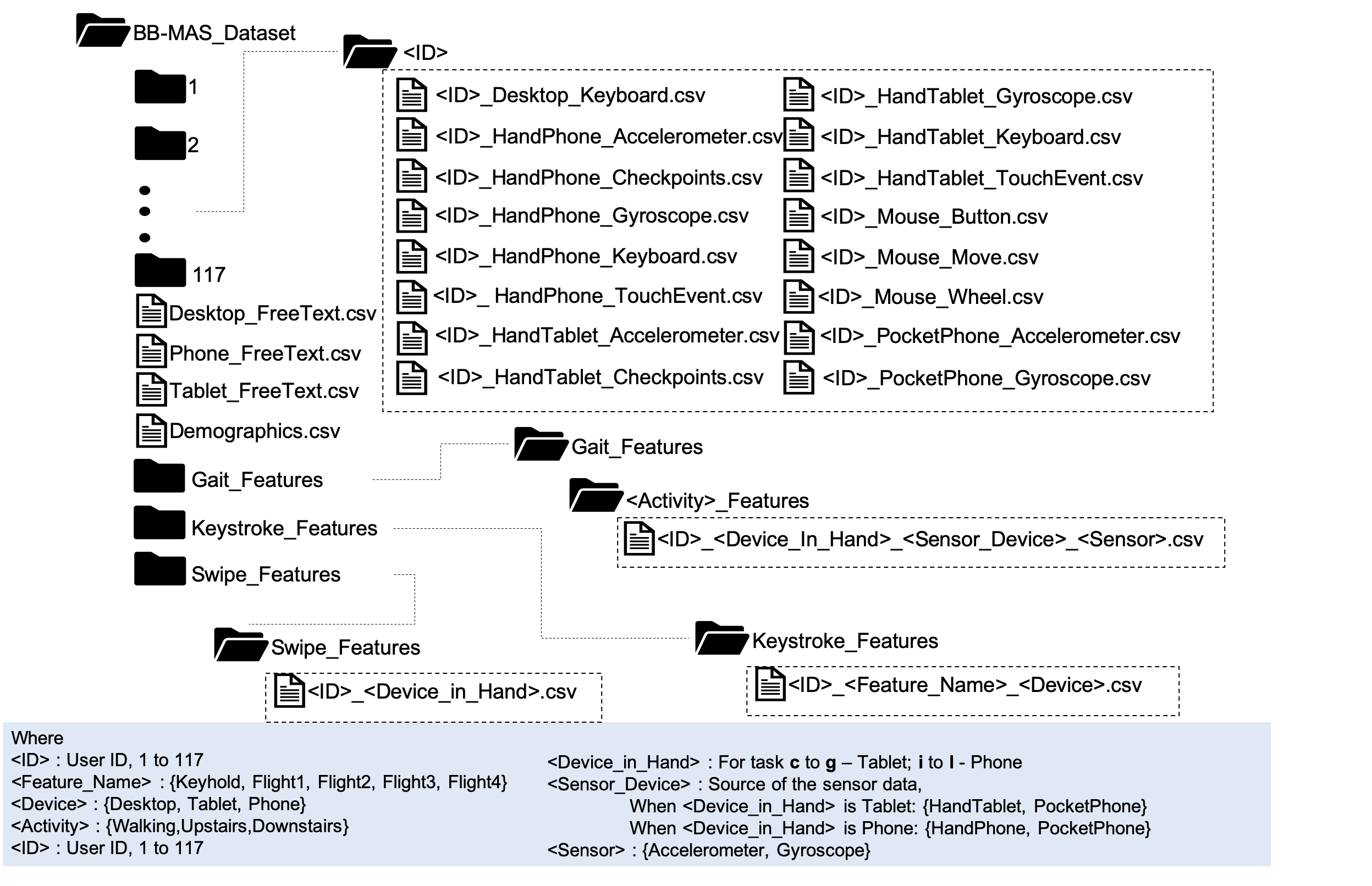}
  \caption{Organisation of the files in our dataset.}~\label{fig:TotalSchema}
\end{figure*}

The raw data from all sensors was originally written to sql databases for speed and accuracy. However, for the convenience of researchers, the raw data and the features extracted from them are organized in simple flat file structure in comma separated format (csv) shared at \textit{http://dx.doi.org/10.21227/rpaz-0h66}. This section elaborates the organization and format of both raw data files and feature extracted files. Fig. \ref{fig:TotalSchema} gives an overview of the entire dataset. It is important that the dataset is clearly understood by its researchers for successful research. Therefore, we explain our dataset in great detail in this section.

\subsubsection{Raw Data}\label{rawdata}
The raw data from each sensor for each user is stored in folder labelled with the user's ID. As shown in Fig. \ref{fig:TotalSchema}, folders "1" to "117" contain the raw data files for each user, the prefix $<$ID$>$ is used to denote the user's ID. The details and format of the raw data files are as follows:
 \begin{itemize}[leftmargin=*]
 \item \textbf{Keystroke Data}:  The temporal data of every key press and release performed by the subject during tasks \textbf{a}, \textbf{b} and \textbf{h} (Table \ref{tab:CollectionTable}) were logged. These files are named;
  \begin{itemize}[leftmargin=*]
  \item[-]$<$ID$>${$\_$}Desktop$\_$Keyboard.csv
  \item[-]$<$ID$>${$\_$}HandTablet$\_$Keyboard.csv
  \item[-]$<$ID$>${$\_$}HandPhone$\_$Keyboard.csv 
  \end{itemize}
  accordingly. These files consist four columns, "EID": event ID (Integer); "key": the key triggering the key-event (String); "direction": the type of key-event (Integer, 0 for press and 1 for release); and "time": the timestamp of the key-event (String in date-time format with millisecond resolution). Table \ref{tab:keyexample} provides an example  of keystroke files with a snippet from user 1 in our dataset.
\begin{table}[]
\caption{Example for keystroke files from user 1.}
\label{tab:keyexample}
\centering
\begin{tabular}{p{0.2cm}lll}
EID & key & direction & time                    \\
0  & t   & 0 & 2017-04-14 18:09:41.538 \\
1  & t   & 1 & 2017-04-14 18:09:41.679 \\
2  & i   & 0 & 2017-04-14 18:09:41.819 \\
.. & .. & .. & ..
\end{tabular}
\end{table}

\item \textbf{Mouse Data}: In addition to keystrokes, data from mouse usage was also collected during task \textbf{a} (Table \ref{tab:CollectionTable}). Please note that there were sampling issues with the mouse data resulting in smaller files, they are included, nonetheless. Mouse events such as, movement, button and wheel were logged into files named; 
\begin{itemize}[leftmargin=*]
  \item[-]$<$ID$>${$\_$}Mouse$\_$Move.csv
  \item[-]$<$ID$>${$\_$}Mouse$\_$Button.csv
  \item[-]$<$ID$>${$\_$}Mouse$\_$Wheel.csv 
  \end{itemize}
  respectively. The Mouse\_Move file has six columns, "EID": event ID (Integer); "rX" and "rY": the x and y coordinates relative to the active window (Integer); "pX" and "pY": the x and y coordinate on screen (Integer); and "time": the timestamp of the mouse-event (String in date-time format with millisecond resolution). The Mouse\_Button file has eight columns, six of them are the same as described for Mouse\_Move, "LR": mouse button (Integer, 0 for left or 1 for right) and "state": type of button event (Integer, 0 for press and 1 for release) are the additional columns. The Mouse\_Wheel file has seven columns, six of them are the same as described for Mouse\_Move in addition to, "delta": direction of scroll (Integer, Negative for scroll-down and positive for scroll-up). Tables \ref{tab:mousemoveexample}, \ref{tab:mousebuttonexample} and \ref{tab:mousewheelexample} provide an example mouse movement, button and wheel data respectively, from user 1.
  
\begin{table}[h]
\caption{Example for mouse movement data from user 1.}
\label{tab:mousemoveexample}
\centering
\begin{tabular}{p{0.2cm}lllll}
EID & rX & rY & pX & pY & time                    \\
0   & 4    & -8   & 1004  & 577   & 2017-04-14 18:09:29.948 \\
1   & 8    & -14  & 1919  & 0     & 2017-04-14 18:09:30.228 \\
2   & -2   & -26  & 1916  & 0     & 2017-04-14 18:21:13.712 \\
.. & .. & .. & .. &.. & .. 
\end{tabular}
\end{table}

\begin{table}[]
\caption{Example for mouse button data from user 1.}
\label{tab:mousebuttonexample}
\centering
\begin{tabular}{p{0.2cm}p{0.3cm}p{0.3cm}p{0.3cm}p{0.2cm}p{0.2cm}p{0.3cm}l}
EID & rX & rY & pX & pY & LR & state & time                    \\
0& 6& -4&   1285&   242&0&  0&  2017-04-14 18:21:17.783 \\
1& -1& 3&   811&    265&0&  1&  2017-04-14 18:21:21.761 \\
2& 0& 0&    811&    265&0&  0&  2017-04-14 18:21:22.120 \\
.. & .. & .. & .. &.. & .. & .. & ..
\end{tabular}
\end{table}

\begin{table}[]
\caption{Example for mouse wheel data from user 1.}
\label{tab:mousewheelexample}
\centering
\begin{tabular}{p{0.2cm}llllll}
EID & rX & rY & pX & pY & delta & time                    \\
0  & 0    & 0    & 1594  & 708   & 120   & 2017-04-14 18:23:10.936 \\
1  & 0    & 0    & 1545  & 708   & 120   & 2017-04-14 18:23:12.000 \\
2  & 0    & 0    & 1618  & 708   & 120   & 2017-04-14 18:23:12.575 \\
.. & .. & .. & .. &.. & .. & .. 
\end{tabular}
\end{table}

\item \textbf{Accelerometer and Gyroscope Data}: For tasks from \textbf{b} through \textbf{m} (Table \ref{tab:CollectionTable}), the values from accelerometer and gyroscope sensors were logged on suitable devices, such as tablet: for tasks \textbf{c} - \textbf{g}; phone in pocket: for tasks \textbf{c} - \textbf{g} and \textbf{i} - \textbf{m}; and phone in hand: for tasks \textbf{i} - \textbf{g}. The sampling rate for these sensors was about 100Hz. The files with accelerometer and gyroscope from the tablet are named;
\begin{itemize}[leftmargin=*]
  \item[-]$<$ID$>${$\_$}HandTablet$\_$Accelerometer.csv
  \item[-]$<$ID$>${$\_$}HandTablet$\_$Gyroscope.csv
  \end{itemize}
those from the phone in the pocket are named;
\begin{itemize}[leftmargin=*]
  \item[-]$<$ID$>${$\_$}PocketPhone$\_$Accelerometer.csv
  \item[-]$<$ID$>${$\_$}PocketPhone$\_$Gyroscope.csv
  \end{itemize}
and from the phone in hand are named;
\begin{itemize}[leftmargin=*]
  \item[-]$<$ID$>${$\_$}HandPhone$\_$Accelerometer.csv
  \item[-]$<$ID$>${$\_$}HandPhone$\_$Gyroscope.csv
  \end{itemize}
  respectively. The accelerometer files have five columns, "EID": event ID (Integer); "Xvalue", "Yvalue", "Zvalue": the acceleration force in m/s$^2$ on x, y and z axes respectively, excluding the force of gravity (Float); and "time": the timestamp of the data point (String in date-time format with millisecond resolution). The gyroscope data have the same five columns, but "Xvalue", "Yvalue" and "Zvalue" is the rate of rotation in rad/s around x, y and z axis respectively (Float). Tables \ref{tab:accelerometerexample} and \ref{tab:gyroscopeexample} show an example for accelerometer and gyroscope data respectively, from user 1.
  
 \begin{table}[]
 \caption{Example for accelerometer data from user 1.}
\label{tab:accelerometerexample}
\centering
\begin{tabular}{lllll}
EID & Xvalue             & Yvalue             & Zvalue            & time                    \\
0  & 1.043 & 3.245 & 9.087  & 2017-04-14 18:56:40.215 \\
1  & 0.995 & 3.303 & 8.936 & 2017-04-14 18:56:40.216 \\
2  & 0.988 & 3.355  & 8.880 & 2017-04-14 18:56:40.234 \\
.. & .. & .. & .. &.. 
\end{tabular}
\end{table}

\begin{table}[]
 \caption{Example for gyroscope data from user 1.}
\label{tab:gyroscopeexample}
\centering
\begin{tabular}{lllll}
EID & Xvalue                & Yvalue               & Zvalue                & time                    \\
0  & -0.045  & 0.036  & -0.013& 2017-04-14 18:56:40.440 \\
1  & -0.027 & 0.027 & -0.017 & 2017-04-14 18:56:40.449 \\
2  & -0.013 & 0.022 & -0.017 & 2017-04-14 18:56:40.461\\
.. & .. & .. & .. &.. 
\end{tabular}
\end{table}

\item \textbf{Swipe Data}: For tasks \textbf{b} and \textbf{h}, data from swipes were recorded on the tablet and phone in hand respectively. These were logged into files named;
\begin{itemize}[leftmargin=*]
  \item[-]$<$ID$>${$\_$}HandTablet$\_$TouchEvent.csv
  \item[-]$<$ID$>${$\_$}HandPhone$\_$TouchEvent.csv
  \end{itemize}
  respectively. The touch data files have ten columns, "EID": event ID (Integer); "Xvalue"  and "Yvalue": the x and y coordinates on screen (Float), "pressure": the approximate pressure applied to the surface by a finger (Float, normalized to a range from 0 (no pressure at all) to 1 (normal pressure)), "touchMajor" and "touchMinor": the length of the major and minor axis, respectively, of an ellipse that represents the touch area (Float, display pixels), "pointerID": index of the pointer/touch used in case of multiple touch points (Integer), "fingerOrientation": the orientation of the finger in radians relative to the vertical plane of the device (Float, 0 radians indicates that the major axis oriented upwards, is perfectly circular or is of unknown orientation), "actionType": indicates the type of event (Integer, 0: finger down/swipe begin, 1: finger up/swipe end and 2:finger move/swipe); and "time": the timestamp of the data point (String in date-time format with millisecond resolution).

\begin{table*}[]
 \caption{Example for swipe data from user 1.}
\label{tab:swipeexample}
\centering
\begin{tabular}{llllllllll}
ID & Xvalue     & Yvalue        & pressure  & touchMajor& touchMinor& pointerID & fingerOrientation & actionType & time \\
0 & 818.085     & 1546.25       & 1.0       & 0.976     & 0.488     & 0         & 0.0               & 0.0 & 2017-4-14 18:56:47:185 \\
1 & 819.140     & 1545.0        & 1.0       & 2.929     & 2.929     & 0         & 0.0               &     & null                   \\
2 & 820.074     & 1543.893      & 1.0       & 2.929     & 2.929     & 0         & 0.0               & 2.0 & 2017-4-14 18:56:47:209
\end{tabular}
\end{table*}

\item \textbf{Checkpoints Data}: For the tasks \textbf{c} - \textbf{g} and \textbf{i} - \textbf{m}, we require checkpoints to separate the data into walking, upstairs and downstairs. The participants were asked to click on buttons on tablet (\textbf{c} - \textbf{g}) or phone in hand (\textbf{i} - \textbf{m}) to mark the opening and closing of doors and start and end of stairs. These checkpoints can be used to separate the data from all other sensors into different activities. Please note that a proctor followed the users during these tasks (making sure not to influence the activity) and noted down incidents where some users clicked the buttons either early or late by a few seconds, adjustments to such timestamps were made manually by adding or subtracting the number of seconds noted down by the proctor. 
Checkpoint files have three columns, "EID":  event ID (Integer); "eventType": type of event (String, \textit{DoorEntry}: user at doorway and is about to open door, \textit{DoorExit}: user has crossed the doorway and the door has closed behind them, \textit{StairEntry}: user about to start climbing up or down the staircase and \textit{StairExit}: user has completed climbing up or down a staircase.); and "time": the timestamp of the data point (String in date-time format with millisecond resolution). Table \ref{tab:checkpointexample} shows an example of checkpoint data from user 1. Using the checkpoint data, the accelerometer and gyroscope data can be segmented into three; a) between "DoorExit" and "DoorEntry" event represents walking on a flat corridor; b) between the first "StairEntry" and "StairExit" represents going downstairs; and between second "StairEntry and "StairExit" represents going upstairs.  

\begin{table}[]
 \caption{Example for checkpoint data from user 1.}
\label{tab:checkpointexample}
\centering
\begin{tabular}{lll}
EID & eventType & time \\
0 & DoorEntry & 2017-04-14 19:41:45.980 \\
1 & DoorExit & 2017-04-14 19:41:50.639 \\
.. & .. & .. \\
4 & StairEntry & 2017-04-14 19:42:18.724\\
5 & StairExit & 2017-04-14 19:42:39.105\\
.. & .. & .. 
\end{tabular}
\end{table}

\item \textbf{FreeText Data}: In tasks \textbf{a}, \textbf{b} and \textbf{h}, the users had to first transcribe two pieces of fixed text;  a) "this is a test to see if the words that i type are unique to me. there are two sentences in this data sample."\footnote{{The transcription sentences were selected based on two criteria: (1) inclusion of many frequently used words in the Oxford English Corpus, and (2) encouraging typing activity on both hands (on both sides on the keyboard). Transcription sentences were typed in lower case.}\label{note1}}; and b) "second session will have different set of lines. carefully selected not to overlap with the first collection phase."\footnotemark[\value{footnote}]. The files labelled "Desktop\_FreeText.csv", "Tablet\_FreeText.csv" and "Phone\_FreeText.csv" provide the timestamp for each user, at which they completed the transcription section  and moved to free text section. In our analyses we have considered the entire typing activity as a whole, these files are provided for researchers who may want to separate  fixed text and free text for their work.

\end{itemize}

\subsubsection{Features from Raw Data}\label{features}
We extracted popular features that are used in literature for each modality. The feature extraction for our dataset can be grouped into three parts, namely keystroke, gait and swipe features. The files consisting the extracted features have also been included in our dataset. We briefly describe the features and their storage below.

\begin{itemize}[leftmargin=*]

\item \textbf{Keystroke Features}:  We select the common twelve unigraphs (single key) and eighteen digraphs (pair of consecutive keys) that occurred the most number of times in all user's keystroke data. The unigraphs are: "BACKSPACE", "SPACE", "a", "e", "h", "i", "l", "n", "r", "S" and "t". The digraphs are: ('BACKSPACE', 'BACKSPACE'), ('SPACE', 'a'), ('SPACE', 'i'), ('SPACE', 's'), ('SPACE', 't'), ('e', 'SPACE'), ('e', 'n'), ('e', 'r'), ('e', 's'), ('n', 'SPACE'), ('o', 'SPACE'), ('o', 'n'), ('r', 'e'), ('s', 'SPACE'), ('s', 'e'), ('t', 'SPACE'), ('t', 'e') and ('t', 'h'). For a unigraph $K_i$ we extract the $Keyhold$ time of the key as a feature:
\begin{itemize}
\setlength\itemsep{0.05em}
\item $Keyhold_{K_i}$ : $K_i Release$ - $K_i Press$
\end{itemize}
For a digraph $K_i$ and $K_{i+1}$ the following temporal features are extracted:
\begin{itemize}
\item $Flight1_{K_iK_{i+1}}$ : $K_{i+1} Press$ - $K_i Release$
\item $Flight2_{K_iK_{i+1}}$ : $K_{i+1}Release$ - $K_i Release$
\item $Flight3_{K_iK_{i+1}}$ : $K_{i+1} Press$ - $K_i Press$
\item $Flight4_{K_iK_{i+1}}$ : $K_{i+1} Release$ - $K_i Press$
\end{itemize}

\begin{figure}
  \centering
  \includegraphics[width=0.8\linewidth]{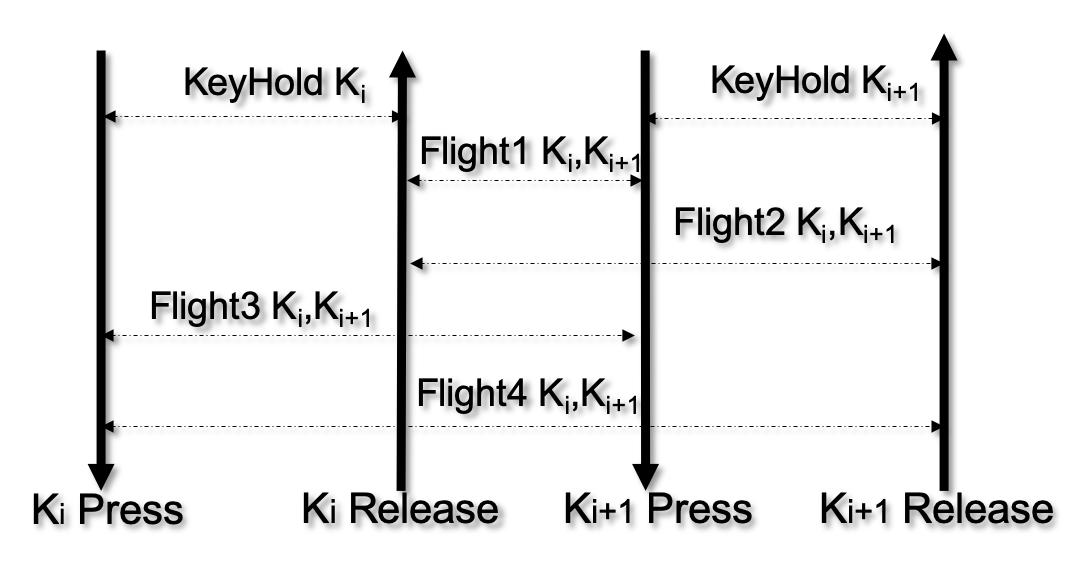}
  \caption{Features extracted from keystroke data.}~\label{fig:keyfea}
\end{figure}
The figure \ref{fig:keyfea} illustrates the temporal features extracted form keystrokes. These files are stored in folder labelled "Keystroke\_Features" which contains the files names with syntax "$<$ID$>$\_$<$Feature\_Name$>$\_$<$Device$>$.csv", where, $<$ID$>$ is the user ID (1-117), $<$Feature\_Name$>$ is the keystroke feature (keyhold, flight1 - flight4) and $<$Device$>$ is either desktop, tablet or phone. Each of these files have column denoting the key ("key" in case of keyhold, "key1" and "key2" in the case of flight) and a column with the value extracted for the feature.

\item \textbf{Gait Features}: As the raw data for the gait is a pair of signals from the accelerometer and gyroscope we extract features from both. The gait data is further subdivided into three activities; "Walking" (on a flat corridor); "Downstairs" (going down the staircase); and "Upstairs" (going up the  staircase). We use a window size of two seconds with a one second overlap between two consecutive windows. For each two second window we extract a host of features from the accelerometer and the gyroscope for x ("Xvalue"), y ("Yvalue"), z ("Zvalue") and m (m=$\sqrt{x^2 + y ^2 + z^2}$). A brief description of the features and their column names in files are as follows: 
\begin{itemize}
\item Mean: mean of x, y, z and m data denoted $xmean$, $ymean$, $zmean$ and $mmean$ respectively.
\item Standard deviation: standard deviation of x, y, z and m data denoted  $xstd$, $ystd$, $zstd$ and $mstd$ respectively.
\item Band power: band power x, y, z and m data denoted $xbp$, $ybp$, $zbp$ and $mbp$  respectively.
\item Energy: energy of the signals x, y, z and m denoted $xenergy$, $yenergy$, $zenergy$ and $menergy$ respectively.
\item Median frequency: median frequency of x, y, z and m signals denoted $xmfreq$, $ymfreq$, $zmfreq$ and $mmfreq$ respectively. 
\item Inter quartile range: the inter quartile range  of x,  y, z and m data denoted $xiqr$, $yiqr$, $ziqr$ and $miqr$ respectively.
\item Range: range of the  x, y, m and z signals denoted $xrange$, $yrange$, $zrange$ and $mrange$ respectively.
\item Signal to noise ratio: the signal to noise ratio in x, y, z and m signals denoted $xsnr$, $ysnr$, $zsnr$  and $msnr$ respectively.
\item Dynamic time warping distance: the DTW distance between pairs of signals x-y, y-z and x-z denoted as $xydtw$, $yzdtw$  and $xzdtw$ respectively. 
\item Mutual information: the mutual information between pairs of signals x-y, x-z, x-m, y-z, y-m and z-m denoted as $xymi$, $xzmi$, $xmmi$, $yzmi$, $ymmi$ and $zmmi$ respectively. 
\item Correlation: the Pearson correlation coefficients between pairs of signals x-y, y-z and x-z signals denoted $xycorr$, $yzcorr$ and $xzcorr$ respectively.  
\end{itemize}
In the "Gait\_Features" folder, we have sub-folders named "$<$Activity$>$\_Features" where activity is either Walking, Downstairs or Upstairs. Each folder consists files with names following the syntax "$<$ID$>$\_$<$Device\_In\_Hand$>$\_ $<$Sensor\_Device$>$\_$<$Sensor$>$.csv", where, $<$ID$>$ is the user ID (1-117), $<$Device\_In\_Hand$>$ is "Tablet" for tasks \textbf{c} - \textbf{g} and Phone for tasks \textbf{i} - \textbf{l}, $<$Sensor\_Device$>$ is the device from which data comes from (HandPhone, HandTablet, PocketPhone) and $<$Sensor$>$ is either accelerometer or gyroscope.

\item \textbf{Swipe Features}: For each swipe performed by users on tablet and phone during tasks \textbf{b} and \textbf{h} respectively, various features related to the speed and trajectory of the swipes are extracted. A brief description of the features and their column names in files are as follows: 
\begin{itemize}
\item Minimum x and y coordinates: the minimum x and y coordinates in the entire swipe denoted by $minx$ and $miny$ respectively.
\item Maximum x and y coordinates: the maximum x and y coordinates in the entire swipe denoted by $maxx$ and $maxy$ respectively.
\item Euclidean distance: the Euclidean distance between the start and end points of the swipe denoted by $eucliddist$.
\item Distance list:  Euclidean distance between points of a swipe denoted by $dlist$.
\item Angle: the tangent angle of the swipe denoted by $tanangle$.
\item Time: the total time taken to for the swipe denoted by $tottime$.
\item Velocity mean and standard deviation: the mean and standard deviation of velocity during the swipe, $vmean$ and $vstd$ respectively.
\item Velocity quartiles: the first, second and third quartiles of velocity during the swipe, $vquarts\_0$, $vquarts\_1$ and $vquarts\_2$ respectively.
\item Acceleration mean and standard deviation: the mean and standard deviation of acceleration during the swipe, $amean$ and $astd$ respectively.
\item Acceleration quartiles: the first, second and third quartiles of acceleration during the swipe, $aquarts\_0$, $aquarts\_1$ and $aquarts\_2$ respectively.
\item Pressure mean and standard deviation: the mean and standard deviation of pressure during the swipe, $pmean$ and $pstd$ respectively.
\item Pressure quartiles: the first, second and third quartiles of pressure during the swipe, $pquarts\_0$, $pquarts\_1$ and $pquarts\_2$ respectively.
\item Area mean and standard deviation: the mean and standard deviation of area during the swipe, $areamean$ and $areastd$ respectively.
\item Area quartiles: the first, second and third quartiles of area during the swipe, $areaquarts\_0$, $areaquarts\_1$ and $areaquarts\_2$ respectively.
\item Direction: the direction of the swipe comparing the displacement of the fingertip in x and y direction, the direction of swipes is deduced as either left, right, up or down  denoted by column $swipetype$.
\end{itemize}
These files are stored in the directory "Swipe\_Features" and named with syntax "$<$ID$>$\_$<$Device\_in\_Hand$>$.csv", where, $<$ID$>$ is the user ID (1-117) and  $<$Device\_in\_Hand$>$ (Tablet or Phone) is the device on which the swipe was performed.
\end{itemize}


\subsection{Demographics}\label{demographics}
Each participant was given a unique ID and made to fill out a brief questionnaire consisting questions relating to demographic, physiology and background. This data is stored in the file labelled "Demographics.csv" in the form of thirteen columns, "User ID": unique ID given to each user (Integer); "Age":  age of the  participant in years (Integer); "Gender": the gender of the participants (Character, "F":  Female, "M": Male  and "O": Other); "Height": height of the participant in inches (Integer, inches]); "Ethnicity": ethnicity of the participant (String); "Languages Spoken": languages that the participant can speak fluently (Tuple, [language1, .., languageN]); "Typing Languages": languages in which the participant can type (Tuple, [language1, .., languageN]); "Handedness": dominant hand for the participant ("Right", "Left" or "Ambidextrous"); "Desktop Hours", "Smartphone  Hours" and "Tablet Hours": approximate range of hours in a day the participant spends using a desktop, phone and tablet respectively (Range, in  hours: 0-1, 2-4, 5-7, 8-12, More than 12); "Typing  Style": denotes touch and  visual typists (Character, "a": Do not look at keyboard/Touch typist, "b": Must look at keyboard/Visual typist and "c": Occasionally look at keyboard/Visual typist); and "Major/Minor": participant's major and minor stream of education (String). 

Table \ref{tab:Collection1demo} summarizes the demographics of the participants in our dataset. The average age of participants in our study was about 25 years with more than half of the participants aged between 23 to 26 years. The youngest and oldest participants were 19 and 35 years of age respectively. The average height of participants was about 67 inches. The shortest and tallest being 54 and 74 inches respectively. The daily usage hours also reflect the popularity of these devices while desktops and phones appear to be used more than tablets. 

\begin{table}[]
 \caption{Summary of demographic data.}
 \label{tab:Collection1demo}
\centering
\begin{tabular}{lll|lll} \hline \hline
& & & & & \\
\multicolumn{2}{c}{Category} & Size & \multicolumn{2}{c}{Category} & Size \\[5pt] \hline 
 & & & & & \\
\multirow{4}{*}{Age in years} & 19 - 22 & 22 & \multirow{5}{*}{\begin{tabular}[c]{@{}l@{}}Daily\\ usage of\\ desktop \\ in hours\end{tabular}} & 0 - 1 & 17 \\ \cline{2-3} \cline{5-6} 
 & & & & & \\
 & 23 - 26 & 61 &  & 2 - 4 & 58 \\ \cline{2-3} \cline{5-6} 
  & & & & & \\
 & 27 - 30 & 28 &  & 5 - 7 & 28 \\\cline{2-3} \cline{5-6} 
  & & & & & \\
 & \textgreater{}30 & 06 &  & 8 - 12 & 12 \\ \cline{1-3} \cline{5-6} 
  & & & & & \\
\multirow{2}{*}{Sex} & Female & 45 &  & \textgreater{}12 & 2 \\ \cline{2-6} 
 & & & & & \\
 & Male & 72 & \multirow{5}{*}{\begin{tabular}[c]{@{}l@{}}Daily\\ usage of \\ phone \\ in hours\end{tabular}} & 0-1 & 3 \\ \cline{1-3} \cline{5-6}
  & & & & & \\
\multirow{4}{*}{\begin{tabular}[c]{@{}l@{}}Height in inches\end{tabular}} & $\leq$60 & 6 &  & 2-4 & 51 \\ \cline{2-3} \cline{5-6} 
 & & & & & \\
 & 60-65 & 40 &  & 5-7 & 43 \\ \cline{2-3} \cline{5-6}
  & & & & & \\
 & 65-70 & 43 &  & 8-12 & 16 \\ \cline{2-3} \cline{5-6} 
  & & & & & \\
 & $>$70 & 28 &  & \textgreater{}12 & 4 \\ \hline
  & & & & & \\
\multirow{5}{*}{\begin{tabular}[c]{@{}l@{}}Spoken\\ Languages\end{tabular}} & 1 & 13 & \multirow{3}{*}{\begin{tabular}[c]{@{}l@{}}Daily\\ usage of \\ tablet\end{tabular}} & 0 - 1 & 93 \\ \cline{2-3} \cline{5-6} 
 & & & & & \\
 & 2 & 64 &  & 2-4 & 20 \\ \cline{2-3} \cline{5-6} 
  & & & & & \\
 & 3 & 32 &  & 5-7 & 4 \\ \cline{2-6} 
  & & & & & \\
 & \multirow{2}{*}{\textgreater{}3} & \multirow{2}{*}{8} & \multirow{2}{*}{\begin{tabular}[c]{@{}l@{}}Typing \\ style\end{tabular}} & Touch & 31 \\ \cline{5-6} 
  & & & & & \\
 &  &  &  & Visual & 86 \\ \hline

\end{tabular}
\end{table}

\subsection{Where is it stored and how to obtain it?}\label{wherestored}

The entire dataset, feature files and demographic file are hosted at IEEE-Dataport \cite{ourdata}. The url of the dataset is  \textit{http://dx.doi.org/10.21227/rpaz-0h66} and is open-access complaint, so it can be downloaded with a free IEEE account.

\section{Analysis}\label{analysis}

Through this paper, we aim to provide in-dept description of our dataset. Therefore, we present results from initial research directions that we have explored. We hope the research community will benefit from the dataset and explore the various other directions of research that cannot be addressed in one single research article. We collect the statistics of the keystroke data and compare the average and standard deviation of the keystroke features (Section \ref{features}) across the three different devices. This helps us provide insights about general typing behavior traits on various devices.

\subsection{Statistics of keystroke data and insights from feature values across devices}\label{statistics}

Table \ref{tab:keydatastats} shows the statistics of keystroke data from our dataset. On an average each participant performed around 11,750, 8,950 and 9,400 keystrokes on desktop, tablet and phone respectively. Even in the minimum condition, each participant has performed 4,350, 4,550 and 5,450 keystrokes on the three devices respectively. When combined, the average keystrokes per user across all three devices is over 30,000 keystrokes and in minimum condition about 19,250.
\begin{table}[]
 \caption{Keystroke data statistics: Number of keystroke events.}
 \label{tab:keydatastats}
\centering
\begin{tabular}{lllll} \hline \hline
&&&&\\
& Desktop & Tablet & Phone     & \begin{tabular}[c]{@{}l@{}}All  \\ Devices\end{tabular} \\ \hline 
& & & & \\
Average & 11760  & 8952  & 9395 & 30153 \\
& & & & \\
Stdev   & 2132   & 1584  & 1472 & 3880 \\
& & & & \\
Min     & 4365   & 4580  & 5463  & 19252\\
& & & & \\
Max     & 18716  & 17029 & 14694 & 41828 \\ \hline
\end{tabular}
\end{table}

\textbf{Outlier Detection for Keystroke Features:} From the keystroke data, we extract the all temporal keystroke features that are popular in literature (See Section \ref{features}). We use a simple filter to remove any instances of keys that were held down for two seconds or more. We also remove instances of the inter-key pauses that are greater than two seconds. We assume that these were caused by pauses, where the user is either thinking or receiving instructions during the data collection. 

\subsubsection{Insights from keystroke feature values across devices}To observe how the keystroke features vary across devices, we calculate the average of the average feature values and average of the standard deviation of the feature values from all users in our dataset. To maintain clarity in presentation we use symbols $d1$ through $d18$ to denote the digraphs ($d1$: ('BACKSPACE', 'BACKSPACE'), $d2$: ('SPACE', 'a'), $d3$: ('SPACE', 'i'), $d4$: ('SPACE', 's'), $d5$: ('SPACE', 't'), $d6$: ('e', 'SPACE'), $d7$: ('e', 'n'), $d8$: ('e', 'r'), $d9$: ('e', 's'), $d10$: ('n', 'SPACE'), $d11$: ('o', 'SPACE'), $d12$: ('o', 'n'), $d13$: ('r', 'e'), $d14$: ('s', 'SPACE'), $d15$: ('s', 'e'), $d16$: ('t', 'SPACE'), $d17$: ('t', 'e') and $d18$: ('t', 'h')). Tables \ref{tab:keyholdstats} to \ref{tab:Fligh4stats} present the values  computed for each feature. 

\textbf{Observations:} We observe that participants have considerably less keyhold times for all keys on the hand-held devices (tablet and phone).  The average of the standard deviation in keyhold time are also very small, less than 50 milliseconds, in all cases except for "backspace". However, it is almost completely the opposite when we consider the flight1 to flight4 features from Tables \ref{tab:Flight1stats} to \ref{tab:Fligh4stats}. We observe that in case of hand-held devices the average feature values are larger than those on desktop. Especially in case of flight1, which is also called the inter-key latency, we can see the values are almost doubled for many digraphs ($d6$, $d10$, $d11$ etc.). It is also worth noticing that both the hand-held devices exhibit similar values in most cases and contrasts, if present, are only with the desktop keystroke features.  

\textbf{Insights:} From our observations it appears that participants in general, take longer time between keys on phones and tablets when compared to a desktop keyboard. However, once the key is pressed the release event occurs much sooner on the phones and tablets implying that smaller amount on time is spent with the finger on the key. We posit that this occurrence maybe a result of lesser number of fingers being in contact with the typing surface on hand-held devices. In most cases participants type on tablets and phones with just their thumbs compared to their usage of many more fingers for the desktop keyboard thus increasing the keyhold time and reducing the inter-key latency on desktop. 

\begin{table}[]
 \caption{Summary of keyhold feature statistics. All values are in milliseconds.}
 \label{tab:keyholdstats}
\centering
\begin{tabular}{l|ll|ll|ll} \hline \hline
& \multicolumn{2}{c|}{Desktop} & \multicolumn{2}{c|}{Tablet} & \multicolumn{2}{c}{Phone} \\
Unigraph       & $\mu$(avg) & $\mu$(std) & $\mu$(avg) & $\mu$(std) & $\mu$(avg) & $\mu$(std)\\\hline
& &&&&&\\
bspace & 168     & 211     & 128     & 175     & 128     & 129     \\
space     & 114     & 57      & 77      & 18      & 89      & 17      \\
a         & 137     & 68      & 98      & 22      & 103     & 19      \\
e         & 123     & 58      & 85      & 20      & 90      & 18      \\
h         & 116     & 53      & 73      & 16      & 81      & 16      \\
i         & 119     & 61      & 71      & 16      & 85      & 17      \\
l         & 102     & 50      & 71      & 16      & 89      & 19      \\
n         & 122     & 63      & 72      & 16      & 83      & 16      \\
o         & 118     & 61      & 72      & 15      & 87      & 16      \\
r         & 129     & 63      & 78      & 19      & 85      & 17      \\
s         & 130     & 60      & 87      & 21      & 94      & 18      \\
t         & 116     & 54      & 76      & 20      & 81      & 16     \\\hline
\end{tabular}
\end{table}

\begin{table}[]
\caption{Summary of Flight1 feature statistics. All values are in milliseconds.}
 \label{tab:Flight1stats}
\centering
\begin{tabular}{l|ll|ll|ll} \hline \hline
& \multicolumn{2}{c|}{Desktop} & \multicolumn{2}{c|}{Tablet} & \multicolumn{2}{c}{Phone} \\
Digraph       & $\mu$(avg) & $\mu$(std) & $\mu$(avg) & $\mu$(std) & $\mu$(avg) & $\mu$(std)\\\hline
& &&&&&\\
$d1$ & 20       & 258     & 80       & 280     & 20      & 228     \\
$d2$             & 205     & 247     & 412     & 318     & 360     & 313     \\
$d3$             & 277     & 247     & 513     & 316     & 472     & 290     \\
$d4$             & 224     & 245     & 432     & 307     & 433     & 321     \\
$d5$             & 218     & 240     & 441     & 327     & 408     & 298     \\
$d6$             & 96      & 167     & 199     & 194     & 162     & 170     \\
$d7$                 & 115     & 151     & 175     & 138     & 144     & 126     \\
$d8$                 & 37      & 110     & 123     & 77      & 130     & 63      \\
$d9$                 & 123     & 114     & 169     & 87      & 162     & 80      \\
$d10$             & 95      & 137     & 200     & 161     & 186     & 142     \\
$d11$             & 95      & 110     & 254     & 151     & 236     & 120     \\
$d12$                 & 99      & 110     & 205     & 94      & 181     & 75      \\
$d13$                 & 22      & 92      & 118     & 73      & 121     & 59      \\
$d14$             & 99      & 163     & 208     & 230     & 163     & 169     \\
$d15$                 & 87      & 97      & 148     & 67      & 147     & 63      \\
$d16$             & 111     & 158     & 250     & 219     & 198     & 154     \\
$d17$                 & 58      & 94      & 138     & 81      & 136     & 68      \\
$d18$                 & 63      & 88      & 111     & 79      & 121     & 74     \\\hline
\end{tabular}
\end{table}

\begin{table}[]
\caption{Summary of Flight2 feature statistics. All values are in milliseconds.}
 \label{tab:Fligh2stats}
\centering
\begin{tabular}{l|ll|ll|ll} \hline \hline

& \multicolumn{2}{c|}{Desktop} & \multicolumn{2}{c|}{Tablet} & \multicolumn{2}{c}{Phone} \\
Digraph       & $\mu$(avg) & $\mu$(std) & $\mu$(avg) & $\mu$(std) & $\mu$(avg) & $\mu$(std)\\ \hline
& &&&&&\\
$d1$ & 166     & 169     & 163     & 154     & 173     & 118     \\
$d2$             & 333     & 248     & 509     & 310     & 462     & 308     \\
$d3$             & 387     & 246     & 582     & 307     & 556     & 287     \\
$d4$             & 348     & 247     & 520     & 301     & 529     & 317     \\
$d5$             & 327     & 238     & 518     & 321     & 488     & 295     \\
$d6$             & 204     & 165     & 278     & 192     & 253     & 170     \\
$d7$                 & 220     & 156     & 250     & 137     & 231     & 126     \\
$d8$                 & 174     & 114     & 200     & 80      & 217     & 65      \\
$d9$                 & 247     & 119     & 255     & 87      & 253     & 82      \\
$d10$             & 212     & 148     & 275     & 157     & 274     & 144     \\
$d11$             & 206     & 117     & 333     & 151     & 326     & 121     \\
$d12$                 & 228     & 122     & 275     & 94      & 263     & 76      \\
$d13$                 & 155     & 93      & 203     & 77      & 209     & 62      \\
$d14$             & 209     & 165     & 286     & 225     & 255     & 170     \\
$d15$                 & 208     & 103     & 230     & 69      & 238     & 65      \\
$d16$             & 219     & 158     & 328     & 215     & 288     & 154     \\
$d17$                 & 181     & 101     & 225     & 82      & 224     & 69      \\
$d18$                 & 182     & 97      & 186     & 79      & 202     & 77     \\\hline
\end{tabular}
\end{table}

\begin{table}[]
\caption{Summary of Flight3 feature statistics. All values are in milliseconds.}
 \label{tab:Fligh3stats}
\centering
\begin{tabular}{l|ll|ll|ll} \hline \hline
& \multicolumn{2}{c|}{Desktop} & \multicolumn{2}{c|}{Tablet} & \multicolumn{2}{c}{Phone} \\
Digraph       & $\mu$(avg) & $\mu$(std) & $\mu$(avg) & $\mu$(std) & $\mu$(avg) & $\mu$(std)\\\hline
& &&&&&\\
$d1$ & 194     & 156     & 179     & 146     & 187     & 110     \\
$d2$       & 314     & 245     & 491     & 311     & 450     & 306     \\
$d3$       & 387     & 248     & 588     & 307     & 559     & 288     \\
$d4$       & 337     & 245     & 509     & 300     & 524     & 315     \\
$d5$       & 326     & 236     & 518     & 320     & 496     & 294     \\
$d6$       & 219     & 167     & 284     & 190     & 252     & 168     \\
$d7$           & 238     & 154     & 265     & 136     & 236     & 120     \\
$d8$           & 179     & 112     & 218     & 77      & 230     & 64      \\
$d9$           & 246     & 122     & 251     & 86      & 248     & 80      \\
$d10$       & 220     & 145     & 267     & 157     & 266     & 143     \\
$d11$       & 210     & 114     & 326     & 152     & 320     & 119     \\
$d12$           & 235     & 121     & 279     & 94      & 268     & 76      \\
$d13$           & 163     & 93      & 197     & 74      & 208     & 60      \\
$d14$       & 229     & 162     & 297     & 224     & 258     & 167     \\
$d15$           & 219     & 100     & 237     & 69      & 246     & 63      \\
$d16$       & 229     & 157     & 324     & 213     & 277     & 151     \\
$d17$           & 178     & 93      & 216     & 81      & 220     & 69      \\
$d18$           & 177     & 90      & 189     & 79      & 203     & 74     \\\hline
\end{tabular}
\end{table}

\begin{table}[]
\caption{Summary of Flight4 feature statistics. All values are in milliseconds.}
 \label{tab:Fligh4stats}
\centering
\begin{tabular}{l|ll|ll|ll} \hline \hline
& \multicolumn{2}{c|}{Desktop} & \multicolumn{2}{c|}{Tablet} & \multicolumn{2}{c}{Phone} \\
Digraph       & $\mu$(avg) & $\mu$(std) & $\mu$(avg) & $\mu$(std) & $\mu$(avg) & $\mu$(std)\\\hline
& &&&&&\\
$d1$ & 380     & 268     & 335     & 230     & 336     & 173     \\
$d2$       & 441     & 249     & 587     & 302     & 552     & 301     \\
$d3$       & 495     & 247     & 656     & 299     & 644     & 285     \\
$d4$       & 459     & 246     & 598     & 294     & 620     & 310     \\
$d5$       & 436     & 239     & 595     & 313     & 577     & 292     \\
$d6$       & 327     & 174     & 363     & 188     & 342     & 167     \\
$d7$           & 343     & 166     & 340     & 135     & 324     & 122     \\
$d8$           & 315     & 132     & 294     & 82      & 317     & 68      \\
$d9$           & 369     & 132     & 337     & 89      & 339     & 82      \\
$d10$       & 336     & 161     & 342     & 155     & 354     & 144     \\
$d11$       & 320     & 123     & 404     & 150     & 410     & 122     \\
$d12$           & 362     & 141     & 350     & 95      & 351     & 78      \\
$d13$           & 296     & 106     & 282     & 79      & 295     & 64      \\
$d14$       & 339     & 172     & 376     & 222     & 348     & 166     \\
$d15$           & 340     & 117     & 319     & 72      & 338     & 67      \\
$d16$       & 337     & 165     & 402     & 210     & 367     & 150     \\
$d17$           & 300     & 107     & 303     & 83      & 307     & 68      \\
$d18$           & 296     & 110     & 264     & 79      & 284     & 77     \\\hline
\end{tabular}
\end{table}

\section{Possible Research Directions}\label{possibledirections}
\begin{itemize}
    \item  \textbf{User authentication for individual devices using keystrokes, gait or swipes:} Our dataset provides multiple modalities, activities and scenarios which can be used separately as individual device or activities for user authentication data. 
    \item \textbf{Activity recognition:} We share data from multiple activities and sub-activities like; free text and fixed text in case of keystrokes; and walking, upstairs and downstairs in case of gait. Recognizing the activity or sub-activity provides better context for methods to be applied.
    \item \textbf{Feature engineering:} Many authentication and identification tasks can be improved with a better understanding of feature sets and their effectiveness for each of the modalities.
    \item \textbf{Inter-Device behavior patterns:} A unique property of our dataset that sets it apart from openly available datasets is that, the same participants performed many overlapping activities on multiple devices. Therefore, inter-device patterns in behavior in same activity or different activities can be researched. For example, \textit{"Can the typing behavior of a user on desktop reveal their typing behavior on phone?".}
    \item \textbf{Physiological or Demographic information leakage in activities:} As we provide a demographic information of each participant, researchers can also explore if the membership of  participants in certain demographic group can be identified from different behavioral activities. 
    \item \textbf{Demographic menagerie:} Existence of groups of users who perform differently at various authentication tasks has been shown in literature \cite{mene1}. Demographic or physiological links to these groupings can be explored. 
\end{itemize}

\section { Comparison with other data sets}\label{comparison}
The datasets that are currently available for behavioral biometrics are collected with the focus on a single activity. We summarize and compare our dataset to other related datasets that are available in literature. We present the key points of comparison with sizeable and related datasets in Tablet \ref{tab:datacomparison}. 
\renewcommand{\arraystretch}{1.5}
\begin{table*}[]
 \caption{Comparison with other related datasets.}
\label{tab:datacomparison}
\centering
\begin{tabular}{lllllll}\hline\hline
                              & Dataset        & \begin{tabular}[c]{@{}l@{}}No. of\\ users\end{tabular} & Type of data          & Type of activity   & \begin{tabular}[c]{@{}l@{}}Device(s)\\ used by\\ participants\end{tabular}                                                           & Highlights                                                                           \\ \hline
\multirow{8}{*}{\rotatebox[origin=c]{90}{Keystroke}} & U@B\_KD \cite{sharedks}       & 148                                                    & Latencies             & Fixed \& Free text & Desktop                                                              & Four different keyboards used \\ \cline{2-7}
                           & SBrook\_KD \cite{prosodyks} & 196                                                    & Latencies             & Free text          & Desktop                                                              & Truthful Vs. Deceptive writing                                                        \\ \cline{2-7}
                           & Video\_KD  \cite{videoks}    & 30                                                     & Latencies \& Video    & Fixed text         & Desktop                                                              & Movement and Motor aspects                                                            \\\cline{2-7}
                           & Pressure\_KD  \cite{pressureks} & 100                                                    & Latencies \& Pressure & Fixed text         & Desktop                                                              & Pressure on desktop keyboard                                                          \\\cline{2-7}
                           & Android\_KD  \cite{androidks}  & 42                                                     & Latencies \& Pressure & Fixed text         & \begin{tabular}[c]{@{}l@{}}37 Tablet \&\\ 5 Phone users\end{tabular} & \begin{tabular}[c]{@{}l@{}}Addition of pressure features\\ for Android\end{tabular}   \\\cline{2-7}
                           & Laser\_2012  \cite{laserks}  & 20                                                     & Latencies             & Fixed \& Free text & Desktop                                                              & Free vs Fixed text behavior                                                       \\   \cline{2-7}
                           & CMU\_KD   \cite{cmuks}     & 51                                                     & Latencies             & Fixed text         & Desktop                                                              & \begin{tabular}[c]{@{}l@{}}Large number of repititions\\ by participants\end{tabular}\\\cline{2-7}
                           & Clarkson\_I \cite{clarksonks}   & 39                                                     & Latencies \& Video    & Fixed \& Free text & Desktop                                                              & \begin{tabular}[c]{@{}l@{}}Video of face and hand\\ while typing\end{tabular} \\  \cline{2-7}   
                           & Clarkson\_II\cite{clarkson2}& 103 & Latencies & Fixed and Free  text & Desktop & \begin{tabular}[c]{@{}l@{}}Natural and uncontrolled\\  includes mouse and app data\end{tabular} \\ \hline  

\multirow{6}{*}{\rotatebox[origin=c]{90}{Gait}}& Kinematics \cite{gait3}  & 42                                                     & \begin{tabular}[c]{@{}l@{}}3d Motion Capture\\ using Force Plates\end{tabular}      & \begin{tabular}[c]{@{}l@{}}Overground \&\\ Treadmill\end{tabular}                                     & \begin{tabular}[c]{@{}l@{}}Force Plates\\ (placed on body)\end{tabular}                                  & \begin{tabular}[c]{@{}l@{}}Anthropometric data \&\\ Pelvis Kinematics\end{tabular}    \\\cline{2-7}
                                                                                         & HuGaDB   \cite{hugadb}   & 18                                                     & \begin{tabular}[c]{@{}l@{}}Accelerometer X 6\\ Gyroscope X 6\\ EMG X 2\end{tabular} & \begin{tabular}[c]{@{}l@{}}12 Activities \\ including Gait,\\ Upstairs and \\ Downstairs\end{tabular} & \begin{tabular}[c]{@{}l@{}}Accelerometer, \\ Gyroscope \&\\ EMG sensors \\ (placed on body)\end{tabular} & \begin{tabular}[c]{@{}l@{}}Body sensor network\\ data\end{tabular}                    \\\cline{2-7}
                                                                                         & VideoGait \cite{gait1}  & 124                                                    & Video                                                                               & Walking                                                                                               & -                                                                                                        & \begin{tabular}[c]{@{}l@{}}Effect of Viewpoint, \\ Clothing \& Carrying\end{tabular}  \\\cline{2-7}
                                                                                         & humanID    \cite{gaitchallenge} & 122                                                    & Video                                                                               & Walking                                                                                               & -                                                                                                        & \begin{tabular}[c]{@{}l@{}}12 Experiments with\\  changes in conditions.\end{tabular} \\\cline{2-7}
                                                                                         & HAR  \cite{har}       & 30                                                     & \begin{tabular}[c]{@{}l@{}}Accelerometer \&\\ Gyroscope\end{tabular}                & \begin{tabular}[c]{@{}l@{}}6 Activities\\ including Gait,\\ Upstairs and \\ Downstairs\end{tabular}   & Smartphone                                                                                               & \begin{tabular}[c]{@{}l@{}}Activity recognition \\ using smartphone\end{tabular}      \\\cline{2-7}
                                                                                         & UniMiB SHAR \cite{unimibshar} & 30                                                     & Accelerometer                                                                       & Gait \& Fall                                                                                          & Smartphone                                                                                               & \begin{tabular}[c]{@{}l@{}}Activity recognition\\ \& fall detection\end{tabular}  \\ \hline   
\multirow{4}{*}{\rotatebox[origin=c]{90}{Swipe}} & ASU\_Touch\cite{touch3}   & 75                                                     & \multirow{4}{*}{\begin{tabular}[c]{@{}l@{}}\\X, Y coordinates\\ Pressure\\ Area of touch\\ Orientation of finger\end{tabular}} & \begin{tabular}[c]{@{}l@{}}Swipe \& Touch\\ Gestures\end{tabular}        & Smartphone                                                                 & \begin{tabular}[c]{@{}l@{}}Re-authentication using\\ swipe gestures\end{tabular}                           \\\cline{2-3}\cline{5-7}
                                                                          & Touchalytics\cite{touchalytics} & 41                                                     &                                                                                                                              & \begin{tabular}[c]{@{}l@{}}Swipe to scroll\\ through images\end{tabular} & Smartphone                                                                 & \begin{tabular}[c]{@{}l@{}}Proposed 30 \\ touch features\end{tabular}                                      \\\cline{2-3}\cline{5-7}
                                                                          & LTU\_Touch\cite{toucheval}   & 190                                                    &                                                                                                                              & \begin{tabular}[c]{@{}l@{}}Swipe through \\ questions\end{tabular}       & Smartphone                                                                 & \begin{tabular}[c]{@{}l@{}}Evaluation of verifiers\\ for touch data\end{tabular}                           \\\cline{2-3}\cline{5-7}
                                                                          & FAST\cite{touchgest}          & 40                                                     &                                                                                                                              & Browsing                                                                 & Smartphone                                                                 & \begin{tabular}[c]{@{}l@{}}Fingergestures \\ Authentication System\\ using Touchscreen (FAST)\end{tabular}\\ \hline

\multirow{3}{*}{\rotatebox[origin=c]{90}{Our Dataset}} & Keystroke & \multirow{3}{*}{\rotatebox[origin=c]{90}{\begin{tabular}[c]{@{}l@{}}Same 117 participants \\ across the datasets\end{tabular}}}& Latencies                                                                                                  & Fixed  and Free text                                                    & \begin{tabular}[c]{@{}l@{}}Desktop, Tablet\\ \& Phone\end{tabular}                       & \multirow{3}{*}{\begin{tabular}[c]{@{}l@{}}\\The same participants\\ performing multiple, common\\ day-to-day activities on multiple\\ devices with real-life placement\\ and usage of devices\end{tabular}} \\ \cline{2-2}\cline{4-6}
                             & Gait      &                                                                                                       & \begin{tabular}[c]{@{}l@{}}Accelerometer X 3\\ Gyroscope X 3\end{tabular}                                  & \begin{tabular}[c]{@{}l@{}}Walking, Upstairs\\ \& Downstairs\end{tabular} & \begin{tabular}[c]{@{}l@{}}Tablet in hand\\ Phone in hand\\ Phone in pocket\end{tabular} &                                                                                                                                                                                                            \\\cline{2-2}\cline{4-6}
                             & Swipe     &                                                                                                       & \begin{tabular}[c]{@{}l@{}}X, Y coordinates\\ Pressure, Area of touch\\ Orientation of finger\end{tabular} & \begin{tabular}[c]{@{}l@{}}Swipe through\\ questions\end{tabular}       & \begin{tabular}[c]{@{}l@{}}Tablet\\ Phone\end{tabular}                                   &                         \\\hline                                                                                                                           
\end{tabular}
\end{table*}

A majority of keystroke datasets are focused on fixed text data with short strings like password, repeated many times by each participant \cite{androidks, pressureks, cmuks}. Most keystroke datasets are collected on desktops \cite{sharedks,prosodyks,cmuks,clarksonks} and very few are on hand-held devices, such as \cite{androidks}, which has a only 42 participants of which 37 participants used tablets and only 5 used phones. Such variations limit the usability of datasets. For gait data, there are two datasets that provide sub-activities, similar to our dataset, such as walking, going up and down the stair case \cite{har,unimibshar}. But, both these datasets have only 30 participants. \textit{HuGaDB} \cite{hugadb}, provides a dataset with 12 different activities collected with a body-sensor-network having 6 accelerometers and gyroscopes each and 2 electromyography (EMG) sensors that are placed on the participants body. Though this data approximates the body movements of participants very closely, it would  not be suitable for continuous authentication due to the unrealistic placement of senors compared to day-to-day use of phones. In case of \textit{VideoGait} and \textit{humanId} (\cite{gait1}, \cite{gaitchallenge}), both having over 120 participants, a third party surveillance approach is more suitable as the  datasets consists of video recordings of participants gait which is not suitable for on-device continuous authentication. 

As touch and swipe as a behavioral biometric is comparatively less explored and the number of datasets is far fewer. All touch and swipe datasets compared in Table \ref{tab:datacomparison} collected similar raw data (coordinates, pressure, area and orientation) while using different ways to make participants perform the swipes. 

All the datasets discussed above provide data for single activity on single device. How a particular activity from a user varies from device to device, or existence of correlation between different activities on different devices cannot be explored with these datasets. Heterogeneity Human-Activity Recognition dataset \cite{hhar} consists data from multiple devices and multiple movement-related activities (no keystrokes), but as data was collected from  only nine users and the device carrying conditions (eight phones carried together in a pouch at the waist and two watches worn on each hand) limit the usability of the dataset for behavioral biometrics.

Therefore, our dataset stands unique by providing data from the same 117 participants performing; a) typing activity, both fixed and free text, on desktop, tablet and phone; b) gait activity, including walking, upstairs and downstairs, with phone in hand, tablet in hand and phone in pocket; and c) swiping activity on tablet and phone.

\section{Lessons learnt}\label{lessons}
The process of collection, curation, pre-processing, storage and sharing of a dataset is indeed challenging. We dealt with various issues at each stage of this effort and share the key lessons learnt, for the benefit of research community, below:
\begin{itemize}[leftmargin=*]
\item Overhead time for the entire process is nontrivial. Time involved in various legalities; preparation and approval of Institutional Review Board (IRB) documents; reaching out and procuring participants; and scheduling them is nontrivial and require great consideration and planning beforehand. 
\item Special attention must be given to avoid Personally Identifiable Information (PII) trickling into data. Especially when data collection aims to capture free text, participants might unknowingly divulge personal information such as names, phone numbers, email ids etc., as part of their answers to questions. This can be corrected either in the protocol designing phase (careful consideration to questions) or in the pre-processing phase.
\item An on-site proctor to oversee each participant's data collection can ensure quality of data especially when data involves capturing key timestamps for activity separation (Section \ref{features}, Checkpoint files). In a few cases we have made manual corrections (by adding or subtracting seconds noted by proctor) to the timestamps where a participant logged them either too early or too late.
\item When data collection involves logging of timestamps on more than one device it is important to make sure the clocks on all devices involved are  synchronized to within a few milliseconds of each other. We carried out several test runs to ensure synchronization of the timestamps on all devices, which was challenging as there were four devices (desktop, pocket-phone, hand-phone and tablet) to be used by every participant. 
\item Before sharing the data publicly, it is important to represent similar data from different devices in the same format for easier usability. For  example; timestamps were logged in a string format (yyyy-mm-dd hr-min-sec.milliseconds) on the desktop and UNIX timestamp on all other devices; and key strokes were logged as characters or keys on the desktop but as ASCII codes on other devices. Therefore, it is better to standardize the data fields before sharing the dataset. 
\item Incomplete data can occur from unexpected application or sensor fault or when participants do not complete the entire process. For example, in our dataset, user 117 did not complete the tasks \textbf{h} to \textbf{m}, but other tasks were complete and are included in the final dataset. In rare cases, where there was too little information for a task or activity, it was better to remove the files for completeness of the shared data. 
\item In a previous data collection effort for keystrokes \cite{prevdata}, we observed some participants tend to fill in low-quality or gibberish text in order to satisfy the minimum text-length criteria (if any) to finish the session earlier. Such occurrences reduce the quality of data and can be remedied either by clearly stating the dos and don'ts to the participants or by the on-site proctor observing and interrupting such behavior.
\end{itemize}

\section{Conclusion and Future Work}\label{conclusion}
Through this paper, we share and provide the details of our large behavioral biometrics dataset for typing, gait and swiping activities of the same user on desktop, tablet and phone (Section \ref{datacollection}). The availability of the data on different devices for the same person makes our dataset unique; and with data from 117 participants, also one of the largest. With this dataset researchers can try to explore questions that were not possible with previously available datasets such as; \textit{"Does the typing of an individual on desktop reveal their typing on a tablet or phone? and vice versa" }; \textit{"Can a person's demographics like age, height, etc., be predicted from the data of typing, gait or swiping activity on any of the devices?"}; to name a few. Each of the files in our dataset are described in detail with example snippets for easier visualization and understanding (Section \ref{rawdata}). 

We explore, describe, extract and analyze the most popular features for each activity in our dataset. All features are described briefly and also included with the dataset repository (Section \ref{features}). The demographics of the participants is shared and includes various physiological and background information with good  spread for most groups (Section \ref{demographics}). The analysis of the features reveals interesting insights. We found participants took  more time between keys on phones and tablets when compared to a desktop keyboard but the release event was much sooner on the phones and tablets implying that smaller amount on time is spent with the finger on the key (Section \ref{analysis}). As the general style of typing on tablets and phones is with just the two thumbs as opposed to several fingers on desktop, we posit that this occurrence maybe a result of lesser number of fingers being in contact with the typing surface on hand-held devices thus increasing the keyhold time and reducing the inter-key latency on desktop. 

This dataset helps address the scarcity in  benchmark datasets for multi-device, multi-activity and multi-modality data from the same participants. Collection of a high-quality dataset that can be publicly shared for the benefit of the community, is indeed a tedious and demanding task. Throughout the process, lessons that we have shared in Section \ref{lessons} are intended for future researchers who make similar endeavors to have an advantage. We also discuss several possible research directions (Section \ref{possibledirections}) that can be explored with the help of this dataset. As part of our future work we will be exploring these directions.


%

\appendices
\section{Cognitive Loads \cite{cog1}}
\label{appendix:cognitive}
\renewcommand{\arraystretch}{1.5}
\begin{tabular}{lcl}
\textbf{Task} & \textbf{Level} & \textbf{Required activity}\\\hline
Remember & 1 & \begin{tabular}{@{}c@{}}Retrieve knowledge from  \\ long-term memory to explain\end{tabular}\\\hline
Understand & 2 & Explain, summarize or interpret\\\hline
Apply & 3 & Apply, execute or implement\\\hline
Analyze & 4 &  \begin{tabular}{@{}c@{}}Organize or break material \\ into constituent parts\end{tabular} \\\hline
Evaluate & 5 & \begin{tabular}{@{}c@{}}Critique or make judgments  \\ based on criteria\end{tabular} \\\hline
Create & 6 & \begin{tabular}{@{}c@{}}Generate, plan or  put elements \\ together\end{tabular}   \\\hline
\end{tabular}


\section{Fixed Text Sentences}
\begin{itemize}
\item  "this is a test to see if the words that i type are unique to me. there are two sentences in this data sample."
\item "second session will have different set of lines. carefully selected not to overlap with the first collection phase."
\end{itemize}

\section{Shopping list}
\begin{itemize}
\item Mountain Bike
\item Plane tickets  from Syracuse, New York to Los Angeles [1 week from today, Coach Seat]
\item Bathing Suit (male or female)
\item Converse All Stair Hiking Boots
\item 24 Pack of Gatorade (24-oz)
\item Ground Transportation (Train, taxi, Bus) from Los Angeles to  San Diego [2 weeks from Today]
\end{itemize}

\section{Free text questions on desktop}
\label{appendix:desktop}
\begin{itemize}
\item List some of the things that you like about Syracuse University.
\item Which internet browser do you typically use (e.g, Google Chrome, Internet Explorer, Mozilla Firefox, etc.)?
\item What improvements would you like to see in that browser?
\item If you were to draw a picture of Syracuse University, what objects would you include in it?
\item What is your favorite vacation spot? Why do you like to visit there?
\item Give step-by-step driving directions to your favorite restaurant in the Syracuse Area, starting from your dorm room/ home.
\item What hobbies or activities are you involved with outside of school/work? Why?
\item Discuss step-by-step instructions for making your favorite type of sandwich. Write them so that the person who has never done this before can follow your instructions.
\item What television programs do you watch for the news and current events? Why? If you do not watch anything on TV, what is your primary source for news information? What do you like about it?
\item Give a brief, but sufficiently detailed plot description of your favorite book, story, or movie.
\item What social networking websites do you  use? What do you like or dislike about these  websites? If you do not use any social network, how do you stay in touch with your friends and acquaintances. Why do you not use social networking websites?
\item Who is your favorite actor, actress, singer, comedian, or TV personality? What do you like about them?

\end{itemize}

\section{Free text and multiple-choice questions on phone}
\label{appendix:phone}
\begin{itemize}
\item What type of Smartphone do you typically use?
    \begin{itemize}
        \item Android
        \item iPhone
        \item Windows
        \item None
        \item Other
    \end{itemize}
\item Which best describes you?
    \begin{itemize}
        \item I have a very active imagination.
        \item I take my civic duties, such as voting, seriously.
        \item I crave excitement.
        \item I would rather cooperate with people than compete with them.
        \item I am a worrier.
        \item I do not like to talk about myself.
    \end{itemize}
\item Of the courses you've taken in college, which was your favorite and why?
\item Think about a class that you did not enjoy. What improvements would you like to see to make the course better?
\item Re-read Question \#2 and the responses. Which response do you feel is least applicable to you and why?
\item Do you intend to pursue an advanced degree (e.g., Master's or Ph.D.)? Why or why not?
\item Find a rule that makes four of the five options alike. Select the option that does not follow this rule:
    \begin{itemize}
        \item 11.28.45.62
        \item 200.217.234.251
        \item 192.209.226.243
        \item 214.231.248.265
        \item 111.127.140.165
     \end{itemize}
\item (Horizontal swipes) Review Question \#7 and the answer that you chose. Why was the rule you found/why did you select your answer?
\item What are the topics of Question \#6 and Question \#10?
\item Give step-by-step directions from this lab space to your dorm room, making specific notes of each time you would descend or ascend stairs.
\end{itemize}

\section{Free text and multiple-choice questions on tablet}
\label{appendix:tablet}
\begin{itemize}
    \item What type of Tablet do you typically use?
     \begin{itemize}
           \item Android
            \item iPad
            \item Windows
            \item Amazon Fire
            \item None
            \item Other
             \end{itemize}
    \item Which best describes you?
     \begin{itemize}
           \item I don't mind bragging about my skills and accomplishments.
           \item I often forget to or neglect to put things back where I found them.
            \item I am dominant, forceful, and/or assertive.
           \item I am easy-going and lackadaisical.
            \item I am set in my ways.
            \item I shy away from crowds.
             \end{itemize}
    \item What is your ideal job after graduation? Why?
    \item Why did you decide to attend Syracuse University?
    \item Re-read Question \#2 and the responses. Which response do you feel is least applicable to you and why?
    \item If all mangoes are golden in color and no golden-colored things are cheap, which of the following is true?
     \begin{itemize}
            \item A. All mangoes are cheap.
           \item B. Golden-colored mangoes are not cheap.
            \item C. Either A or B are true.
            \item D. Both A and B are true.
            \item E. Neither A or B are true.
        \end{itemize}
    \item Review Question \#6 and the answer that you chose. Why did you select your answer?
    \item (Horizontal swipes) If Question \#6 was changed to read "If some mangoes are golden in color and no golden-colored things are cheap", which answer would be correct and why?
    \item What are your thoughts on the current U.S. president? Which policies, if any, would you like to see changed and how?
    \item Discuss step-by-step the process for sending an email from your Syracuse email account. Write these instructions such that a person who has never done this before can follow your instructions.
    \item Please provide any comments that you have about the survey or the experiment thus far.
\end{itemize}



\ifCLASSOPTIONcaptionsoff
  \newpage
\fi



\bibliographystyle{IEEEtran}
\bibliography{IEEE_data.bib}

\begin{thebibliography}{10}
\providecommand{\url}[1]{#1}
\csname url@samestyle\endcsname
\providecommand{\newblock}{\relax}
\providecommand{\bibinfo}[2]{#2}
\providecommand{\BIBentrySTDinterwordspacing}{\spaceskip=0pt\relax}
\providecommand{\BIBentryALTinterwordstretchfactor}{4}
\providecommand{\BIBentryALTinterwordspacing}{\spaceskip=\fontdimen2\font plus
\BIBentryALTinterwordstretchfactor\fontdimen3\font minus
  \fontdimen4\font\relax}
\providecommand{\BIBforeignlanguage}[2]{{%
\expandafter\ifx\csname l@#1\endcsname\relax
\typeout{** WARNING: IEEEtran.bst: No hyphenation pattern has been}%
\typeout{** loaded for the language `#1'. Using the pattern for}%
\typeout{** the default language instead.}%
\else
\language=\csname l@#1\endcsname
\fi
#2}}
\providecommand{\BIBdecl}{\relax}
\BIBdecl

\bibitem{disckeyfea}
K.~S. Balagani, V.~V. Phoha, A.~Ray, and S.~Phoha, ``On the discriminability of
  keystroke feature vectors used in fixed text keystroke authentication,''
  \emph{Pattern Recognition Letters}, vol.~32, no.~7, pp. 1070 -- 1080, 2011.

\bibitem{keystrokemon}
F.~Monrose and A.~D. Rubin, ``Keystroke dynamics as a biometric for
  authentication,'' \emph{Future Generation computer systems}, vol.~16, no.~4,
  pp. 351--359, 2000.

\bibitem{scankeystroke}
A.~{Serwadda}, Z.~{Wang}, P.~{Koch}, S.~{Govindarajan}, R.~{Pokala},
  A.~{Goodkind}, D.~{Brizan}, A.~{Rosenberg}, V.~V. {Phoha}, and K.~{Balagani},
  ``Scan-based evaluation of continuous keystroke authentication systems,''
  \emph{IT Professional}, vol.~15, no.~4, pp. 20--23, July 2013.

\bibitem{biogait}
D.~Gafurov, K.~Helkala, and T.~S{\o}ndrol, ``Biometric gait authentication
  using accelerometer sensor.'' \emph{JCP}, vol.~1, no.~7, pp. 51--59, 2006.

\bibitem{gaitwearable}
D.~Gafurov, E.~Snekkenes, and P.~Bours, ``Gait authentication and
  identification using wearable accelerometer sensor,'' in \emph{2007 IEEE
  workshop on automatic identification advanced technologies}.\hskip 1em plus
  0.5em minus 0.4em\relax IEEE, 2007, pp. 220--225.

\bibitem{gaitsurvey}
\BIBentryALTinterwordspacing
C.~Wan, L.~Wang, and V.~V. Phoha, ``A survey on gait recognition,'' \emph{{ACM}
  Comput. Surv.}, vol.~51, no.~5, pp. 89:1--89:35, 2019. [Online]. Available:
  \url{https://dl.acm.org/citation.cfm?id=3230633}
\BIBentrySTDinterwordspacing

\bibitem{touchgest}
T.~{Feng}, Z.~{Liu}, K.~{Kwon}, W.~{Shi}, B.~{Carbunar}, Y.~{Jiang}, and
  N.~{Nguyen}, ``Continuous mobile authentication using touchscreen gestures,''
  in \emph{2012 IEEE Conference on Technologies for Homeland Security (HST)},
  Nov 2012, pp. 451--456.

\bibitem{swipegest}
S.~{Mondal} and P.~{Bours}, ``Swipe gesture based continuous authentication for
  mobile devices,'' in \emph{2015 International Conference on Biometrics
  (ICB)}, May 2015, pp. 458--465.

\bibitem{toucheval}
\BIBentryALTinterwordspacing
A.~Serwadda, V.~V. Phoha, and Z.~Wang, ``Which verifiers work?: {A} benchmark
  evaluation of touch-based authentication algorithms,'' in \emph{{IEEE} Sixth
  International Conference on Biometrics: Theory, Applications and Systems,
  {BTAS} 2013, Arlington, VA, USA, September 29 - October 2, 2013}, 2013, pp.
  1--8. [Online]. Available: \url{https://doi.org/10.1109/BTAS.2013.6712758}
\BIBentrySTDinterwordspacing

\bibitem{cmuks}
K.~S. Killourhy and R.~A. Maxion, ``Comparing anomaly-detection algorithms for
  keystroke dynamics,'' in \emph{2009 IEEE/IFIP International Conference on
  Dependable Systems \& Networks}.\hskip 1em plus 0.5em minus 0.4em\relax IEEE,
  2009, pp. 125--134.

\bibitem{sharedks}
Y.~{Sun}, H.~{Ceker}, and S.~{Upadhyaya}, ``Shared keystroke dataset for
  continuous authentication,'' in \emph{2016 IEEE International Workshop on
  Information Forensics and Security (WIFS)}, Dec 2016, pp. 1--6.

\bibitem{laserks}
K.~S. Killourhy and R.~A. Maxion, ``Free vs. transcribed text for
  keystroke-dynamics evaluations,'' in \emph{Proceedings of the 2012 Workshop
  on Learning from Authoritative Security Experiment Results}.\hskip 1em plus
  0.5em minus 0.4em\relax ACM, 2012, pp. 1--8.

\bibitem{androidks}
M.~Antal, L.~Z. Szab{\'o}, and I.~L{\'a}szl{\'o}, ``Keystroke dynamics on
  android platform,'' \emph{Procedia Technology}, vol.~19, pp. 820--826, 2015.

\bibitem{pressureks}
C.~C. Loy, C.~P. Lim, and W.~K. Lai, ``Pressure-based typing biometrics user
  authentication using the fuzzy artmap neural network,'' 2005.

\bibitem{prosodyks}
\BIBentryALTinterwordspacing
R.~Banerjee, S.~Feng, J.~S. Kang, and Y.~Choi, ``Keystroke patterns as prosody
  in digital writings: A case study with deceptive reviews and essays,'' in
  \emph{Proceedings of the 2014 Conference on Empirical Methods in Natural
  Language Processing (EMNLP)}.\hskip 1em plus 0.5em minus 0.4em\relax Doha,
  Qatar: Association for Computational Linguistics, October 2014, pp.
  1469--1473. [Online]. Available:
  \url{http://www.aclweb.org/anthologo/D14-1155}
\BIBentrySTDinterwordspacing

\bibitem{videoks}
\BIBentryALTinterwordspacing
A.~M. Feit, D.~Weir, and A.~Oulasvirta, ``How we type: Movement strategies and
  performance in everyday typing,'' in \emph{Proceedings of the 2016 CHI
  Conference on Human Factors in Computing Systems}, ser. CHI '16.\hskip 1em
  plus 0.5em minus 0.4em\relax New York, NY, USA: ACM, 2016, pp. 4262--4273.
  [Online]. Available: \url{http://doi.acm.org/10.1145/2858036.2858233}
\BIBentrySTDinterwordspacing

\bibitem{gait1}
\BIBentryALTinterwordspacing
S.~Yu, D.~Tan, and T.~Tan, ``A framework for evaluating the effect of view
  angle, clothing and carrying condition on gait recognition,'' in
  \emph{Proceedings of the 18th International Conference on Pattern Recognition
  - Volume 04}, ser. ICPR '06.\hskip 1em plus 0.5em minus 0.4em\relax
  Washington, DC, USA: IEEE Computer Society, 2006, pp. 441--444. [Online].
  Available: \url{https://doi.org/10.1109/ICPR.2006.67}
\BIBentrySTDinterwordspacing

\bibitem{gait2}
T.~N. Nguyen, H.~H. Huynh, and J.~Meunier, ``3d reconstruction with
  time-of-flight depth camera and multiple mirrors,'' \emph{IEEE Access},
  vol.~6, pp. 38\,106--38\,114, 2018.

\bibitem{gait3}
\BIBentryALTinterwordspacing
C.~A. Fukuchi, R.~K. Fukuchi, and M.~Duarte, ``A public dataset of overground
  and treadmill walking kinematics and kinetics in healthy individuals,''
  \emph{PeerJ}, vol.~6, pp. e4640--e4640, Apr 2018. [Online]. Available:
  \url{https://www.ncbi.nlm.nih.gov/pubmed/29707431}
\BIBentrySTDinterwordspacing

\bibitem{hugadb}
\BIBentryALTinterwordspacing
R.~Chereshnev and A.~Kert{\'{e}}sz{-}Farkas, ``Hugadb: Human gait database for
  activity recognition from wearable inertial sensor networks,'' \emph{CoRR},
  vol. abs/1705.08506, 2017. [Online]. Available:
  \url{http://arxiv.org/abs/1705.08506}
\BIBentrySTDinterwordspacing

\bibitem{touchalytics}
\BIBentryALTinterwordspacing
M.~Frank, R.~Biedert, E.~Ma, I.~Martinovic, and D.~Song, ``Touchalytics: On the
  applicability of touchscreen input as a behavioral biometric for continuous
  authentication,'' \emph{Trans. Info. For. Sec.}, vol.~8, no.~1, pp. 136--148,
  Jan. 2013. [Online]. Available:
  \url{http://dx.doi.org/10.1109/TIFS.2012.2225048}
\BIBentrySTDinterwordspacing

\bibitem{touch3}
\BIBentryALTinterwordspacing
L.~Li, X.~Zhao, and G.~Xue, ``Unobservable re-authentication for smartphones,''
  in \emph{20th Annual Network and Distributed System Security Symposium,
  {NDSS} 2013, San Diego, California, USA, February 24-27, 2013}, 2013.
  [Online]. Available:
  \url{https://www.ndss-symposium.org/ndss2013/unobservable-re-authentication-smartphones}
\BIBentrySTDinterwordspacing

\bibitem{unimibshar}
\BIBentryALTinterwordspacing
D.~Micucci, M.~Mobilio, and P.~Napoletano, ``Unimib shar: A dataset for human
  activity recognition using acceleration data from smartphones,''
  \emph{Applied Sciences}, vol.~7, no.~10, 2017. [Online]. Available:
  \url{http://www.mdpi.com/2076-3417/7/10/1101}
\BIBentrySTDinterwordspacing

\bibitem{har}
D.~Anguita, A.~Ghio, L.~Oneto, X.~Parra, and J.~L. Reyes-Ortiz, ``Human
  activity recognition on smartphones using a multiclass hardware-friendly
  support vector machine,'' in \emph{Proceedings of the 4th International
  Conference on Ambient Assisted Living and Home Care}, ser. IWAAL'12.\hskip
  1em plus 0.5em minus 0.4em\relax Berlin, Heidelberg: Springer-Verlag, 2012,
  pp. 216--223.

\bibitem{ourdata}
\BIBentryALTinterwordspacing
A.~K.~Belman, L.~Wang, S.~S. Iyengar, P.~Sniatala, R.~Wright, R.~Dora,
  J.~Balwdin, Z.~Jin, and V.~V. Phoha, ``Su-ais bb-mas (syracuse university and
  assured information security - behavioral biometrics multi-device and
  multi-activity data from same users) dataset,'' 2019. [Online]. Available:
  \url{http://dx.doi.org/10.21227/rpaz-0h66}
\BIBentrySTDinterwordspacing

\bibitem{mene1}
N.~{Yager} and T.~{Dunstone}, ``The biometric menagerie,'' \emph{IEEE
  Transactions on Pattern Analysis and Machine Intelligence}, vol.~32, no.~2,
  pp. 220--230, Feb 2010.

\bibitem{clarksonks}
E.~Vural, J.~Huang, D.~Hou, and S.~Schuckers, ``Shared research dataset to
  support development of keystroke authentication,'' in \emph{IEEE
  International Joint Conference on Biometrics}.\hskip 1em plus 0.5em minus
  0.4em\relax IEEE, pp. 1--8.

\bibitem{clarkson2}
C.~{Murphy}, J.~{Huang}, D.~{Hou}, and S.~{Schuckers}, ``Shared dataset on
  natural human-computer interaction to support continuous authentication
  research,'' in \emph{2017 IEEE International Joint Conference on Biometrics
  (IJCB)}, Oct 2017, pp. 525--530.

\bibitem{gaitchallenge}
S.~{Sarkar}, P.~J. {Phillips}, Z.~{Liu}, I.~R. {Vega}, P.~{Grother}, and K.~W.
  {Bowyer}, ``The humanid gait challenge problem: data sets, performance, and
  analysis,'' \emph{IEEE Transactions on Pattern Analysis and Machine
  Intelligence}, vol.~27, no.~2, pp. 162--177, Feb 2005.

\bibitem{hhar}
\BIBentryALTinterwordspacing
A.~Stisen, H.~Blunck, S.~Bhattacharya, T.~S. Prentow, M.~B. Kj{\ae}rgaard,
  A.~Dey, T.~Sonne, and M.~M. Jensen, ``Smart devices are different: Assessing
  and mitigatingmobile sensing heterogeneities for activity recognition,'' in
  \emph{Proceedings of the 13th ACM Conference on Embedded Networked Sensor
  Systems}, ser. SenSys '15.\hskip 1em plus 0.5em minus 0.4em\relax New York,
  NY, USA: ACM, 2015, pp. 127--140. [Online]. Available:
  \url{http://doi.acm.org/10.1145/2809695.2809718}
\BIBentrySTDinterwordspacing

\bibitem{prevdata}
\BIBentryALTinterwordspacing
A.~Serwadda, V.~V. Phoha, and A.~Kiremire, ``Using global knowledge of users'
  typing traits to attack keystroke biometrics templates,'' in
  \emph{Proceedings of the Thirteenth ACM Multimedia Workshop on Multimedia and
  Security}, ser. MM\&\#38;Sec '11.\hskip 1em plus 0.5em minus 0.4em\relax New
  York, NY, USA: ACM, 2011, pp. 51--60. [Online]. Available:
  \url{http://doi.acm.org/10.1145/2037252.2037263}
\BIBentrySTDinterwordspacing

\bibitem{cog1}
D.~G. Brizan, A.~Goodkind, P.~Koch, K.~Balagani, V.~V. Phoha, and A.~Rosenberg,
  ``Utilizing linguistically enhanced keystroke dynamics to predict typist
  cognition and demographics,'' \emph{International Journal of Human-Computer
  Studies}, vol.~82, pp. 57 -- 68, 2015.

\end{thebibliography}
%

%








\end{document}